\documentclass[aip,reprint,pop]{revtex4-1}

\usepackage[colorlinks=true,allcolors=blue]{hyperref}
\usepackage{bm}
\usepackage{amssymb}
\usepackage{lipsum}
\usepackage{graphicx}
\usepackage{dcolumn}
\usepackage{verbatim}
\usepackage{amsmath}
\usepackage[T1]{fontenc}
\usepackage{mathptmx}
\usepackage{float}

\expandafter\ifx\csname package@font\endcsname\relax\else
\expandafter\expandafter
\expandafter\usepackage
\expandafter\expandafter
\expandafter{\csname package@font\endcsname}%
\fi
\begin{document}

\title{Experimental investigation of coaxial-gun-formed plasmas injected into a background transverse magnetic field or plasma}

\author{Yue Zhang}
\thanks{Invited speaker. Present address: Univ. of Washington, Seattle, WA 98195}
\affiliation{University of New Mexico, Albuquerque, New Mexico, 87131 USA}

\author{Dustin M. Fisher}
\affiliation{University of New Mexico, Albuquerque, New Mexico, 87131 USA}

\author{Mark Gilmore}
\email{mgilmore@unm.edu}
\affiliation{University of New Mexico, Albuquerque, New Mexico, 87131 USA}

\author{Scott C. Hsu}
\affiliation{Los Alamos National Laboratory, Los Alamos, New Mexico, 87545 USA}

\author{Alan G. Lynn}
\thanks{Present address: Naval Research Laboratory, Washington, DC 20375}
\affiliation{University of New Mexico, Albuquerque, New Mexico, 87131 USA}

\date{\today}
\begin{abstract}
Injection of coaxial-gun-formed magnetized plasmas into a background transverse vacuum magnetic field or into a background magnetized plasma has been studied in the helicon-cathode (HelCat) linear plasma device at the University of New Mexico [M. Gilmore et al., J. Plasma Phys.~\textbf{81}, 345810104 (2015)]. A magnetized plasma jet launched into a background transverse magnetic field shows emergent kink stabilization of the jet due to the formation of a sheared flow in the jet above the kink stabilization threshold $0.1 k V_A$ [Y. Zhang et al., Phys.\ Plasmas {\bf 24}, 110702 (2017)]. Injection of a spheromak-like plasma into a transverse background magnetic field led to the observation of finger-like structures on the side with a stronger magnetic field null between the spheromak and background field. The finger-like structures are consistent with magneto-Rayleigh-Taylor instability. Jets or spheromaks launched into a background, low-$\beta$ magnetized plasma show similar behavior as above, respectively, in both cases.
\end{abstract}
\maketitle

\section{Introduction}
The interaction of a dynamic plasma with a background magnetic field or magnetized plasma is a fundamental process relevant to a broad range of plasma physics.~Considerable efforts have been devoted to the study of this interaction such as the simulation of astrophysical jets,\cite{honda2002self} bipolar outflows associated with young stellar object (YSO),\cite{pelletier1992hydromagnetic} active galactic
nuclei (AGN),\cite{clausen2011signatures} solar-wind evolution,\cite{schatten1969model,jokipii1990polar} and magnetotail physics,\cite{zmuda1970characteristics,lyons2013quantitative} etc.~In magnetic fusion, cross-magnetic-field injection to deliver fuel into the core of a tokamak is critical for more efficient utilization of deuterium-tritium fuel.\cite{leonard1986trapping,brown1990spheromak,brown1990current,raman1994experimental,loewenhardt1995performance,raman1997experimental,mclean1998design,olynyk2008development} Additional interest in this stems from attempts to investigate the magnetohydrodynamic (MHD) instability of the plasma boundary for optimizing the energy confinement time for high fusion gain.\cite{ebrahimi2016dynamo,ebrahimi2017nonlinear} The experimental study of plasma motion across a transverse magnetic field has a long history.\cite{tuck1959plasma,schmidt1960plasma,baker1965experimental,parks1988refueling} Modern experimental investigations include those using pulsed-power-driven plasma guns,\cite{hsu2002laboratory,hsu2003experimental,hsu2005jets,bellan2005simulating,yun2007large} radial wire array Z pinches,\cite{lebedev2002laboratory,lebedev2004jet,lebedev2005magnetic,ciardi2007evolution}
and laser-produced plasmas.\cite{mostovych1989laser,ripin1990laboratory,harilal2004confinement} Most of these basic experimental studies have been of plasma launched into vacuum, except for fueling studies. \cite{brown1990current,brown1990spheromak,matsumoto2016characterization} In addition, nonlinear ideal MHD simulations have modeled a plasma injected into a magnetized background plasma under varying conditions.\cite{liu2008ideal,liu2009ideal,liu2011ideal}

In this paper, we present the Plasma Bubble Expansion eXperiment (PBEX) results of the interaction of a dynamic argon plasma with a uniform background transverse magnetic field.~We explore this interaction evolution with multiple diagnostics.~The magnetic tension force from the rising curvature of the background transverse magnetic field is crucial for observed features reported here.~Comparison with existing theories indicates that the observations are consistent with expected advection, stabilization, transition, and instability growth times. 

For PBEX, first, a magnetized plasma jet is launched into a background transverse magnetic field. A more stable jet is observed compared to launching into vacuum. By measuring plasma density, temperature, velocity distributions and global magnetic field configuration, the possible mechanisms responsible for jet stabilization are isolated and the most likely candidate for the observed plasma jet dynamics is verified. We propose that the background magnetic tension force, due to the increase of curvature of the background magnetic field, compresses the jet as the ambient magnetic field increases along the jet path and causes a sheared axial flow gradient in the jet body. This emergent axial sheared flow appears to provide the kink stabilization mechanism.\cite{zhang2017emergent} Measurements of the shear-flow gradient are consistent with axial shear stabilization theory.\cite{shu1995sheared}

Second, a spheromak-like plasma (plasma bubble) is launched into a background transverse magnetic field. The tension force slows down the bubble propagation speed, leading to the observation of finger-like structures on the side with the stronger field null between the spheromak and the background field. The observations are consistent with magneto-Rayleigh-Taylor (mRT) instability.

Experiments have also been carried out in which coaxial-gun-formed plasmas are injected into a background plasma (temperature $T_{e} \sim 1-5 eV$, density $n_{e} \sim 10^{17}~m^{-3}$) that has been characterized elsewhere.\cite{desjardins2016dynamics,kelly2016ari} The background plasma pressure is typically 10$-$100 times less than the background magnetic pressure (i.e., low $\beta \sim 0.01-0.1$). Therefore, the background magnetic field dominates the interaction, and the plasma itself has a relatively small effect on the observations reported in this paper.

Some of the results in this paper are closely related to prior work.~For example, our observations of various plasma morphologies and the development of an $n=1$ kink instability when our gun-formed plasmas are injected into a background field or plasma are very similar to prior work that studied plasma injection into a vacuum.\cite{yee2000taylor,hsu2002laboratory,hsu2003experimental,hsu2005jets} However, our observation of kink stabilization due to emergent sheared flow arising from the interaction of the injected plasma with the background field or magnetized plasma is a new result.\cite{zhang2017emergent} In addition, our observation of finger-like structures when a spheromak is injected into a background field or magnetized plasma is reminiscent of the mRT instability observed due to an instability cascade resulting from a disrupting kink.\cite{moser2012magnetic} However, in our case, the finger-like structures, which are shown to be consistent with mRT, arise due to deceleration of the spheromak as it expands into a background field or magnetized plasma. Localization of the observed mRT structures coincide with the region of greatest field cancellation, which minimizes the magnetic stabilization term of the mRT.

The paper is organized as follows.~In Sec.~II, the PBEX apparatus, diagnostics, control and acquisition systems are discussed in detail.~In Sec.~III, the experimental results are presented for launching plasma jets and bubbles into a background transverse magnetic field, along with detailed physics interpretations.~Sec.~IV presents the experimental results for coaxial-gun-formed plasmas injected into a background magnetized plasma.~A summary is given in Sec.~V.
\begin{figure*}[htb]
\includegraphics[width=\textwidth]{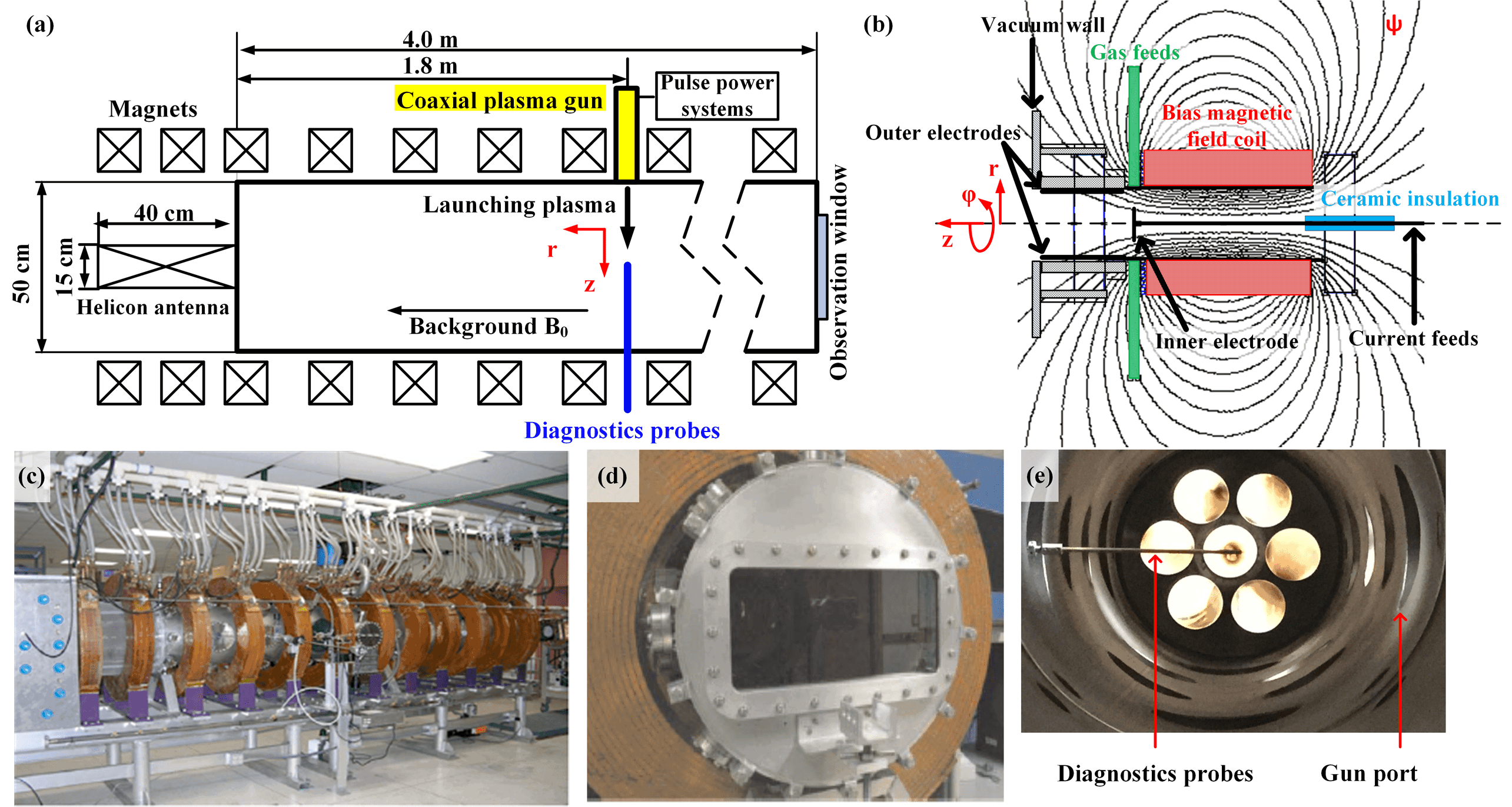}
\caption{(Color online). (a) Top-view schematic of HelCat vacuum chamber and coaxial gun setup, not to scale. Magnetic and electrostatic probes measure localized plasma parameters, and an intensified CCD camera detects coaxial-gun-formed plasmas evolution via an end-view window. (b) Side-view schematic of coaxial gun, showing the inner and outer electrodes (black), gas feed line (green), bias magnetic coil (red), current-feed-through (blue), contours of constant bias poloidal flux, and cylindrical coordinate system. Details are discussed at Sec. II. B. (c) Picture of HelCat linear device which provides a background magnetic field and a background helicon plasma. (d) Picture of the end-view window. (e) CCD camera view via the end-view window with gun port and diagnostic probe marked.}
\label{fig:overview}
\end{figure*}

\section{Experimental Apparatus and diagnostics}
The compact coaxial plasma gun was designed to be geometrically as simple as possible: a copper disk works as a cathode, surrounded by a co-axial cylindrical annulus as an anode. The gun is mounted on a side port 1.8 m away from the helicon source, with an angle of $90^{\circ}$ with respect to the background magnetic field, as indicated in Fig.~\ref{fig:overview} (a). A bias magnetic field coil is located on the cylindrical body, as shown in in Fig.~\ref{fig:overview} (b). The various components of the experimental setup are described below.

\subsection{HelCat linear device}
HelCat, shown in Fig.~\ref{fig:overview} (a) and (c), is an approximately 4-m long, 50-cm diameter cylindrical vacuum chamber with multiple 10-inch conflat type ports.~The chamber has a large rectangular window (39-cm~$\times$~18.5-cm) at the cathode-source-end that provides a good view of the entire cross-section of the vacuum chamber, as shown in Fig.~\ref{fig:overview} (d) and (e). Vacuum is maintained at $5\times10^{-7}$ Torr by a 1000 L/min turbo molecular pump at the mid-chamber with gate valve and backing systems. Background magnetic field is produced by thirteen sets of water-cooled solenoidal magnetic coils along the chamber axis as shown in Fig.~\ref{fig:overview} (c). The currents flowing through these coils provide various background magnetic field conditions for PBEX. For instance, a 500-G steady state magnetic field is produced with a direct current of 114 A. Magnetic field ripple is kept to $<$1$\%$ along the chamber axis and $\sim3\%$ at the plasma edge near $r = 20$ cm. Further details of the HelCat linear device are reported elsewhere.\cite{lynn2009helcat,gilmore2015helcat}
\subsection{Compact magnetized coaxial gun}
The compact magnetized coaxial plasma gun, \cite{zhang2009design} shown in Fig.~\ref{fig:overview} (b), is a 10-cm long cylinder and consists of (1) an inner copper tube (0.009 cm wall) with a 2.54 cm diameter copper disk at the left-most end as an inner electrode, (2) a cylindrical coaxial 5.08 cm diameter copper annulus as an outer electrode, and (3) an external solenoidal coil that covers the main cylindrical body of the gun, providing the bias magnetic flux. The coil consists of 4 layers of 110 turns each layer (440 turns total) of 12 American wire gauge (AWG) insulated square magnet wire, and has an inductance of 6.25 mH and resistance of 0.5 $\Omega$. A 1.27 cm width annulus vacuum gap separates the inner and outer electrodes. There are four 0.64 cm diameter gas puff feed lines symmetrically located around the gun to achieve a fast and localized gas fill in the electrode gap. Details of the gas injection system are given in Sec. II. D.

The current feed-through is mounted on a 3-3/$8^{\prime\prime}$ conflat flange.~The inner electrode extends 10.16 cm from the flange face on the air side and 12.65 cm on the vacuum side. The insulator of the inner electrode is an alumina ceramic tube, holding up to 12 kV. The outer cylindrical electrode is mounted via stainless steel bolts directly to the vacuum chamber, electrically connected to the vacuum chamber ground.

The experiments are characterized using the cylindrical coordinate system shown in Fig.~\ref{fig:overview} (b): the $\phi$ and (z-r) direction will be referred to as toroidal and poloidal, respectively. Contours of constant poloidal flux generated by the bias coil are also shown in Fig.~\ref{fig:overview} (b).
\subsection{Power systems}
\begin{figure}
\includegraphics[width=0.5\textwidth]{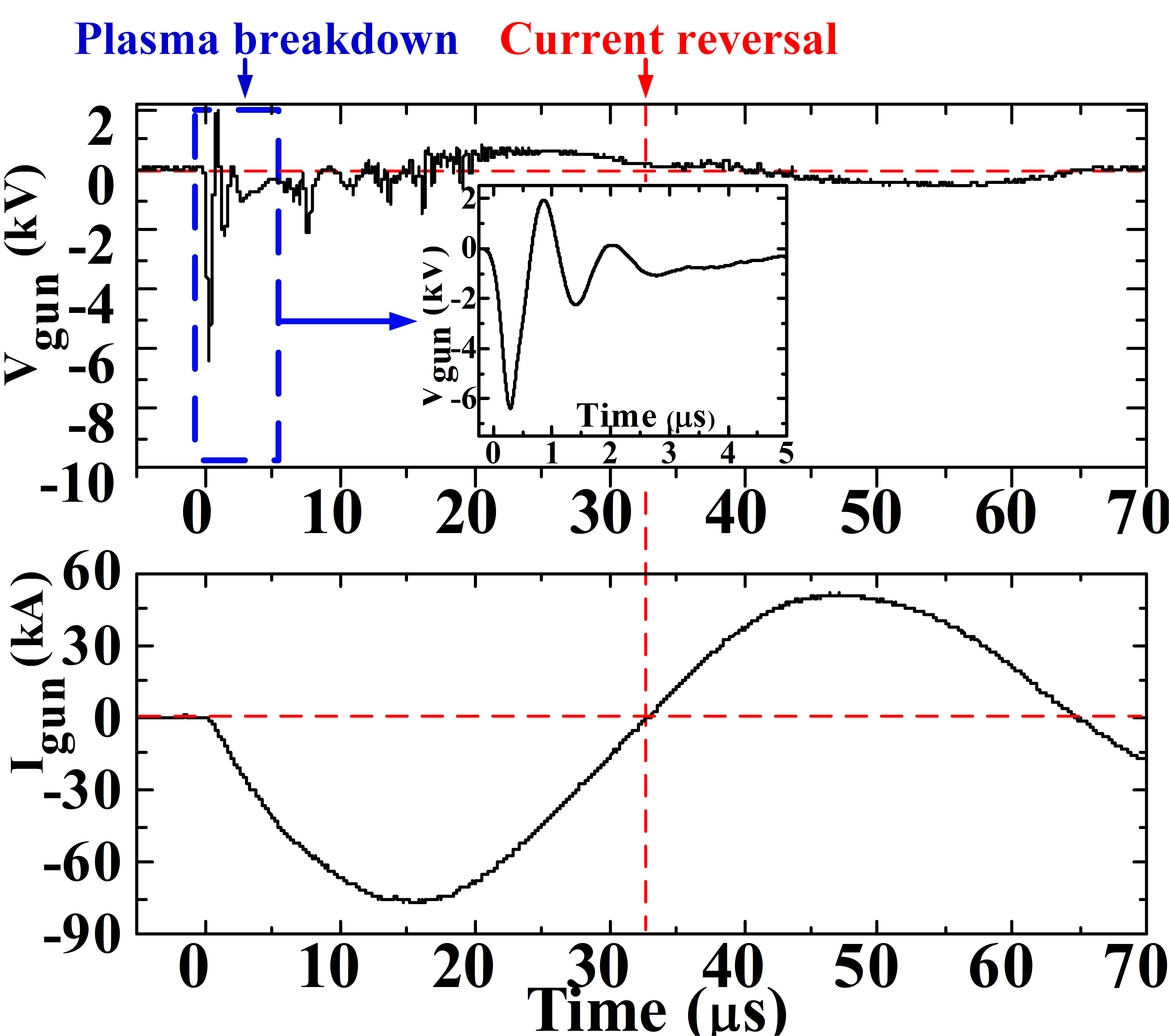}
\caption{Typical $V_{gun}$ (top) and $I_{gun}$ (bottom) traces. Initial charge voltage is 6 kV. Breakdown occurs at t$\sim$2$\mu$s. Peak current is 60 kA.}
\label{fig:typicalIV}
\end{figure}

The coaxial plasma gun is powered by an ignitron-switched 120 $\mu$F main capacitor bank (cap-bank). The main cap-bank is connected to the gun electrodes via eight, parallel, double-layer, low inductance (0.51 $\mu$H/m) coaxial cables (Belden YK-198), which carry electrical power from the main cap-bank to the gun electrodes with low line loss. Typical operation at a cap-bank voltage of 6 kV yields the discharge peak current, $I_{gun}$, $\sim$ 60 kA, and gun voltage, $V_{gun}$, $\approx$ 1 kV (after breakdown). Characteristic traces of $I_{gun}$ and $V_{gun}$ are shown in Fig.~\ref{fig:typicalIV}

The external bias magnetic field coil is powered by a cap-bank array of total 60 mF. The half period of the discharge current trace for the external coil is $90~ms$, reaching peak at $\sim 23 ~ms$ and is effectively constant on the $10~\mu$s time scale of PBEX. The bias poloidal magnetic flux created by the external bias coil, $\psi_{gun}$, is typically 0.3-1.3 mWb.
\subsection{Gas injection}
A fast gas puff system is important for PBEX since Paschen breakdown and efficient gas ionization need a highly localized gas pressure on the order of 100 mTorr. The fast gas valve, utilized in PBEX, provides a localized high pressure natural argon gas in the gap between the inner and outer electrodes. The basic operating principle is that via discharging a pulse current in a single-turn coil, the gas valve induces an image current in an adjacent disk.~The disk is thus repelled from the coil, generating a transient natural gas flow from the gas feed lines into the gap space between electrodes.~The restoring force on the disk is supplied by a combination of high pressure gas from the feed line and a mechanism spring, the details of which are reported elsewhere.\cite{thomas1993simple} Optimum timing of gas valve firing is determined empirically by adjusting the gas valve triggering time to minimize the time delay between the main cap-bank trigger and gas breakdown, typically a few microseconds as indicated in Fig.~\ref{fig:typicalIV} (top). One gas valve is employed for a total of four gas injection holes. A typical operation voltage is 1kV and calibrations indicate that each pulse injects $10^{20}-10^{21}$ argon molecules, 3.6$\times10^{-6}$~g in argon mass.
\subsection{Diagnostics}
Multiple diagnostics are employed in PBEX to study the dynamic interaction between the coaxial-gun-formed plasmas and background transverse magnetic field/magnetized plasma. The details of the diagnostics are discussed here.

The images presented in this paper are taken with a multiple-frame charge-coupled device (CCD) camera, Ultra UHSi 12-24.\cite{UHSi1224} The specifications for this camera are follows: 12 frames per shot, 1000$\times$860 pixels per image, 12 bits per pixel, and a 5 $ns$ minimum exposure time. The camera is positioned in front of the cathode-source-end vacuum chamber window shown in Fig.~\ref{fig:overview} (d).~The typical camera settings for PBEX are: 12 frames with 1 $\mu$s exposure time at framing rate 500k frames per second (fps) per plasma gun shot ($2~\mu$s interframe time). The images are recorded by a professional computer (PC) via Ethernet and post-pulse analyzed using the software IVV Imprint.

Magnetic field data are taken with a radial array of commercial pickup coils on a stainless-steel shaft. Thirty-three commercial chip inductors are grouped into 11 clusters, glued to 11 notch channels of a
Delrin probe holder. In each channel cell, three chip inductors are held together to measure three components of magnetic field in cartesian coordinates. The spacing between each notch is 8.55 mm, thus the effective probe length is 108.55 mm. Also, by placing the B-dot probe at different space positions along the coaxial-gun-formed plasma propagation axis, the plasma propagation velocities at multiple radii and positions are estimated under different experimental settings.

Furthermore, plasma density and temperature are measured by a multiple-tip Langmuir probe.~A Pearson coil and an attenuating high voltage probe are utilized to measure the discharge current, $I_{gun}$, and voltage, $V_{gun}$.
\subsection{Control and data acquisition}
From the discussion above, several various pulsed power systems (bias coil, gas valve, and main cap-bank for the coaxial gun) need to be triggered in a proper time sequence for the plasma gun operation, as well for the data acquisition system to digitize and record diagnostic signals. The digital delay generator (California Avionics Laboratories, INC. Model No.~IO3CR) is employed to provide multiple controllable trigger signals for various pulse power systems. A typical trigger sequence is as follows: $\left(1\right)$ the bias coil is fired at $t = 0~ms$, $\left(2\right)$ the gas valve is fired at $t = 10~ms$ (experimentally decided, accommodating the neutral gas travel time through the gas feed line into the plasma formation region), $\left(3\right)$ the main cap-bank ignitron-switch is trigged at $t=23~ms$ (considering the $23~ms$ rising time of the bias coil power system to reach the maximum bias flux) with gas breakdown typically occurring 1$-$2 $\mu$s later. At the same time ($t = 23~ms$), the digitized data acquisition system is triggered, starting to record diagnostic probes data.

The data acquisition system is primarily made of three analog digitizers (Joerger Enterprises, INC. Model No. TR). Each digitizer has 16 individual channels with sampling rate of 40 MHz, 12-bit resolution per channel, 100 k$\Omega$ input impedance, 512 kB memory, and an input range of -5 V$\sim$+5 V with adjustable offset. All magnetic probe and gun diagnostic signals are digitized by this system. The digital acquisition system is controlled and communicated with PC via LabView. Data is analyzed using MATLAB. The magnetic probe array signals are integrated numerically.
\subsection{Plasma parameters}
The plasma produced by the compact coaxial gun have the following global parameters: density $n_{e}\sim10^{20}$ $m^{-3}$,~$T_{e}\sim$ $T_{i}\sim10~eV$, and $B$ $\sim$ 0.2$-$1 kG. The plasma characteristic length scale $l$ is $\sim$ 10 cm. The characteristic Alfv\'en transit time is $\sim$ 2.0 $\mu$s. The plasma lifetime is about 20 $\mu$s. The resistive diffusion time is on the order of 1 ms. The Lundquist number $S\approx 10^2$. Thus, the 20 $\mu$s lifetime lasts several Alfv\'en times, and the magnetic flux is reasonably frozen into the plasma.~For the background helicon plasma, the basic plasma parameters, in comparison with gun-formed-plasma parameters are listed in Table~\ref{tab:table1}. For such a dynamic system, the gun-formed-plasma characteristics exhibit very robust, reproducible measurements for a wide parameter regime. Each experimental setup has explored hundreds of pulses which show similar plasma dynamics.~The pulse-to-pulse standard deviation in the magnetic-field measurements over a large pulses samples is about 15\%.
\begin{table}
\renewcommand\arraystretch{1.5}
\caption{\label{tab:table1}HelCat and gun-plasma parameters: plasma density, ion temperature, electron temperature, magnetic field strength, and ratio $\beta$ of plasma thermal pressure to magnetic pressure.}
\begin{ruledtabular}
\begin{tabular}{ccc}
Parameter & HelCat  & Coaxial gun \\
\hline
$n_{0} (m^{-3})$  & $10^{17}$ & $10^{20}$   \\
$T_{i}$ (eV)      & 0.1    &10              \\
$T_{e}$ (eV)    & 1-5     & 10             \\
$B_{0}$ (G)       & 0-500   & $200-10^{3}$       \\
$\beta$         & 0.01    & 0.1-0.2          \\
\end{tabular}
\end{ruledtabular}
\end{table}

\section{Experimental results}
\subsection{Plasma morphologies resulting from \textbf{$\lambda_{gun}$} parameter scan}
\begin{figure}
\includegraphics[width=0.5\textwidth]{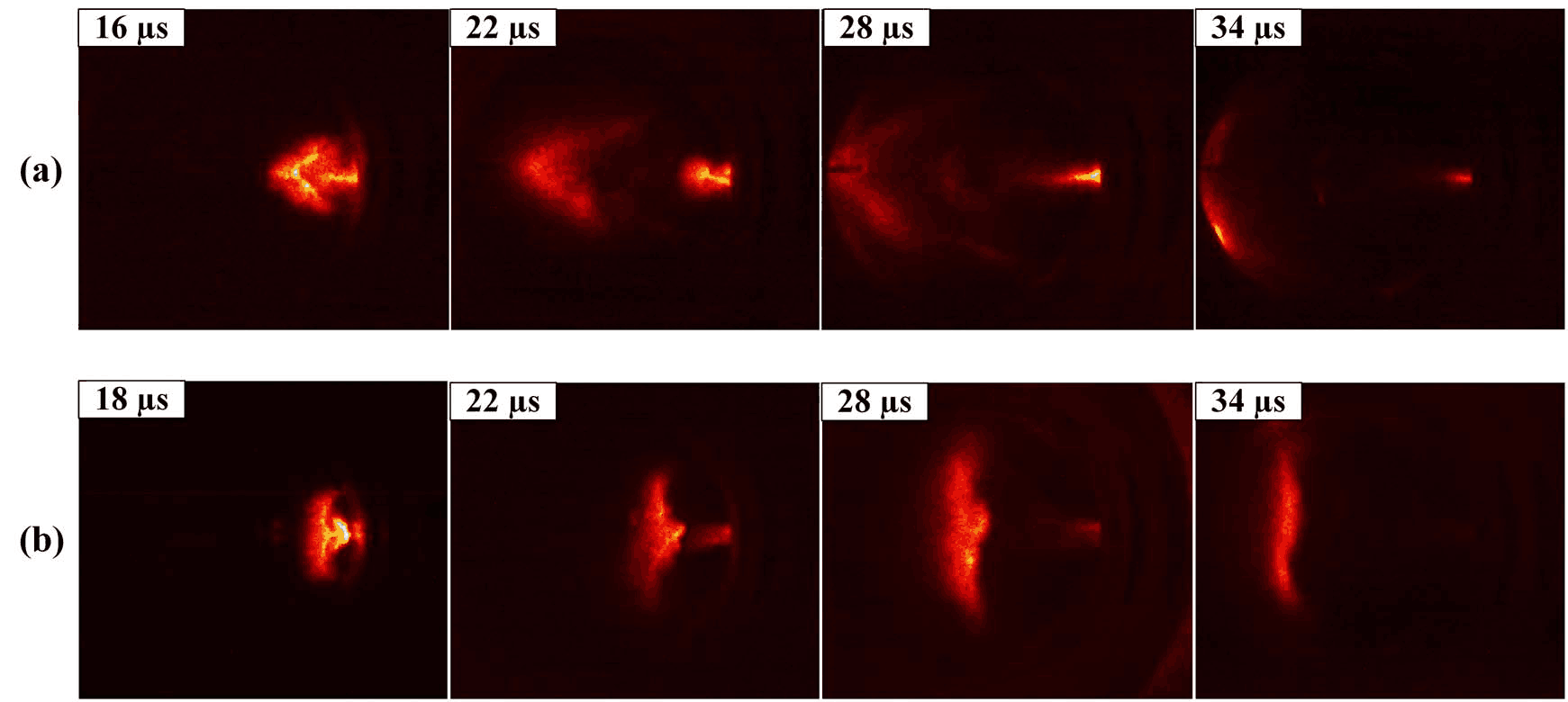}
\caption{(Color online).~CCD images of plasmas evolution: (a) typical plasma jet formation: a typical conical shape leading edge, followed by the plasma jet body itself. With the jet length increasing, $n$=1 kink develops and the jet body breaks; (b) typical plasma spheromak-like (bubble) formation: a symmetrical leading line with clearly scene of detachment from the gun port.}
\label{fig:formations}
\end{figure}

Prior experiments by Yee\cite{yee2000taylor} and Hsu\cite{hsu2002laboratory,hsu2005jets} have shown that coaxial-gun-formed plasmas can evolve into multiple distinct morphology regions depending on the peak $\lambda_{gun}$ ($\lambda_{gun}\equiv\mu_{0}I_{gun}/\psi_{gun}$). The parameter $\lambda_{gun}$ can be controlled experimentally by adjusting the main cap-bank voltage (which controls $I_{gun}$) and the bias coil bank voltage (which controls $\psi_{gun}$). With the image data from CCD camera, two different observed plasma morphologies have been identified: jet and spheromak-like formations, as shown in Fig.~\ref{fig:formations}. False color is applied to the images for ease of viewing. The characteristic features of these two distinct formations and the experimental results of launching these two type of plasma into the background transverse magnetic field will be discussed in detail in this section.

\subsection{Plasma jet injection into background transverse magnetic field}
\subsubsection{Jet kink instability}
\begin{figure}
\includegraphics[width=0.5\textwidth]{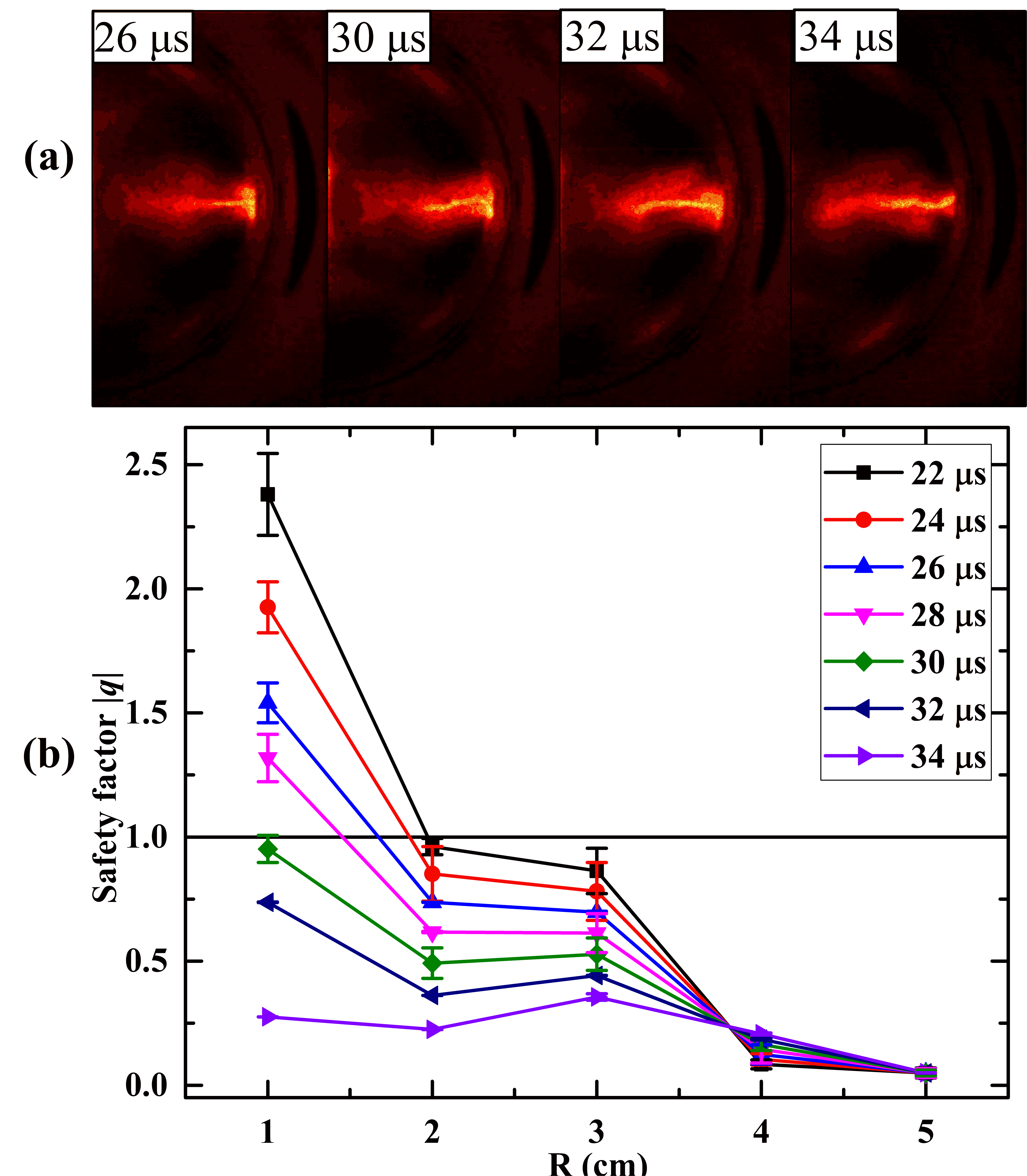}
\caption{(Color online).~(a) $n=1$ kink instability develops as the jet length increases, and (b) safety factor $q$ time evolution at multiple radii. At t~$\sim$~28~$\mu$s, $q$ (R = 1 cm) drops below unity and the jet develops kinks, corresponding to CCD camera images. (Pulse No.\ 001050714)}
\label{fig:jetq}
\end{figure}

Before launching a plasma jet into a background transverse magnetic field, the characteristic features of a plasma jet propagating into the vacuum chamber were investigated. For a current-driven plasma jet, as the length increases, the classical $n=1$ kink instability starts to develop in the jet column as shown in Fig.~\ref{fig:jetq} (a). Prior works have addressed the jet $n=1$ kink instability.\cite{hsu2002laboratory,hsu2003experimental,hsu2005jets} Compared with the Kruskal-Shafranov theory of current-driven instability, the instability parameter, safety factor $q$ and condition for instability is given by
\begin{equation}
\label{eq:1}
q=\frac{2\pi aB_{z}}{lB_{\varphi }}<1,
\end{equation}
where $\mathit{a}$ and $\mathit{l}$ are the jet radius and length, respectively.

The developed kinks in the PBEX plasma jet have been analyzed and examined using Eq.~\ref{eq:1}. Fig.~\ref{fig:jetq} shows images of the evolution as well as the calculated safety factor $q$ from the measured poloidal and toroidal magnetic field data, and the estimated jet length and radius from the CCD images of plasma jet emission.~Typically, the plasma is observed in unfiltered visible light.~Although the exact relationship between emission intensity and plasma density is complex, we use the light emission (allowing for a generous uncertainty) as a proxy for inferring the radius and length of the bulk mass of the jet.~We allow for a conservative $\pm 50$\% error bar in the inferred radius and length to calculate the safety factor $q$ as shown in Fig.~\ref{fig:jetq} (b). During the jet evolution, at early times, $q$ has a value greater than 1 and the associated jet is stable. As the jet progresses in time, its length increases and $q$ decreases. At t~$\sim$~28~$\mu$s, $q$ drops below unity and the jet develops kinks.

\begin{figure}
\includegraphics[width=0.5\textwidth]{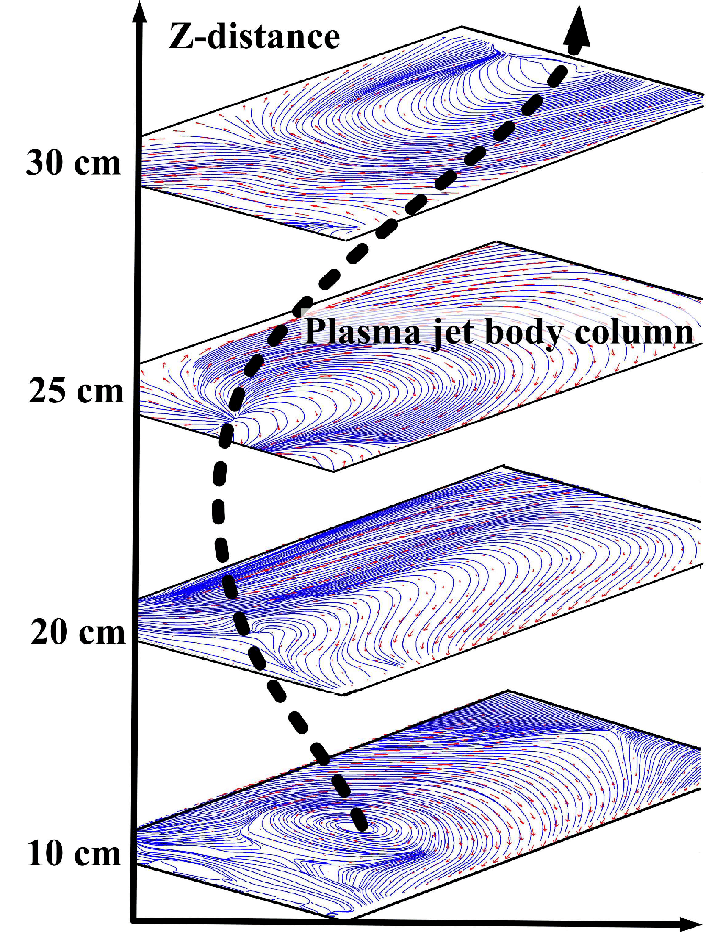}
\caption{(Color online).~Toroidal magnetic field contour plots at multiple Z locations indicates $n = 1$ kink instability develops in jet body column (dash line). Data are taken from Pulse No.\ 013042314, No.\ 043050714, No.\ 044050714, No.\ 012050914, No.\ 013050914 and No.\ 022050914.}
\label{fig:toroidalcontour}
\end{figure}

Furthermore, by placing the B-dot probe array at multiple positions along the jet propagation axis (+z direction), the direct measured toroidal magnetic fields contour plots also indicate the developed kink instability in the jet body column (assuming the jet body is located at the center of concentric streamlines) as shown in Fig.~\ref{fig:toroidalcontour}.

\subsubsection{A more stable jet is observed}
\begin{figure}
\includegraphics[width=0.5\textwidth]{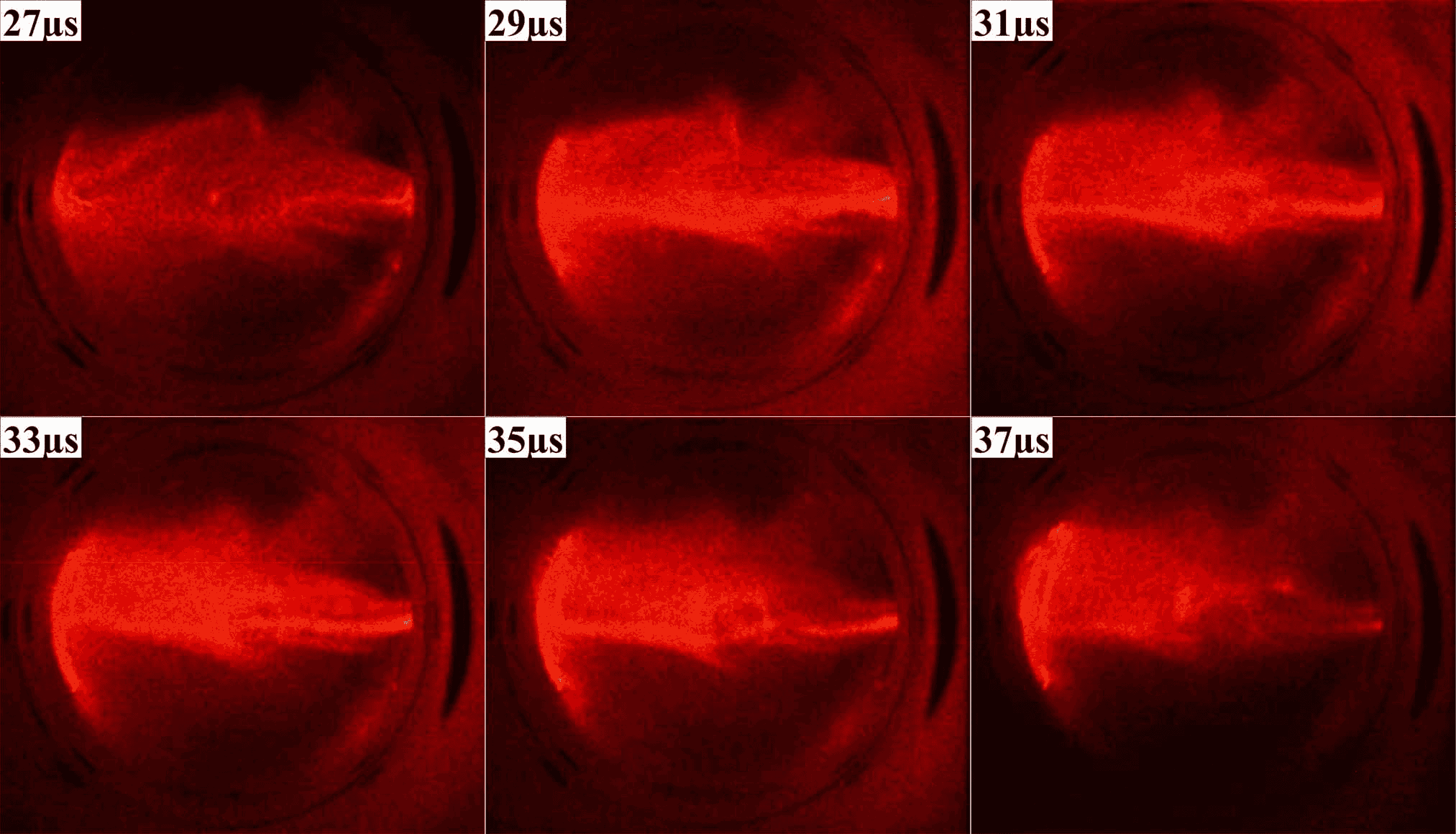}
\caption{(Color online).~A more-stable plasma jet column is observed when injecting into a 500-G magnetic field transverse to the direction of jet propagation.(pulse No.\ 008050114, 2-$\mu$s interframe time)}
\label{fig:bgjet}
\end{figure}

Under the same operation settings, the plasma jet is launched into a 500-G background magnetic field (transverse to the direction of jet propagation). The jet evolution for this case is shown in Fig.~\ref{fig:bgjet}.~The images indicate that (1) the plasma jet penetrates the background magnetic field and (2) a more stable jet is formed during the injection. As inferred from light emission in Fig.~\ref{fig:bgjet}, the jet length is on the order of 50 cm, which is much longer than the approximately 10-cm length observed for the vacuum case in Fig.~\ref{fig:jetq} (a). The jet life time is $\sim3t_{Alfv\acute{\mathrm{e}}n}$~(where~$t_{Alfv\acute{\mathrm{e}}n}\equiv l/V_{Alfv\acute{\mathrm{e}}n}$), which is prolonged compared to the vacuum case for which the plasma-jet lifetime is $\sim$ $t_{Alfv\acute{\mathrm{e}}n}$. Comparing the experimental settings for these two cases, the only difference is the presence of the background magnetic field in the main vacuum chamber.

High magnetic Reynolds number indicates that the plasma jet excludes the background transverse magnetic field from its interior as it penetrates.~The background magnetic field is then advected with the plasma jet propagation and induces a magnetic tension force to act on the plasma jet. Such a magnetic field penetration mechanism has been discussed by numerous investigators.\cite{chapman1931magn,tuck1959plasma,baker1965experimental} We propose that the background magnetic tension force causes a sheared axial flow gradient in the jet body, which appears to provide the kink stabilization mechanism. The details of the experimental results are reported elsewhere.\cite{zhang2017emergent}
\begin{figure}[htb]
\centering
\includegraphics[width=0.45\textwidth]{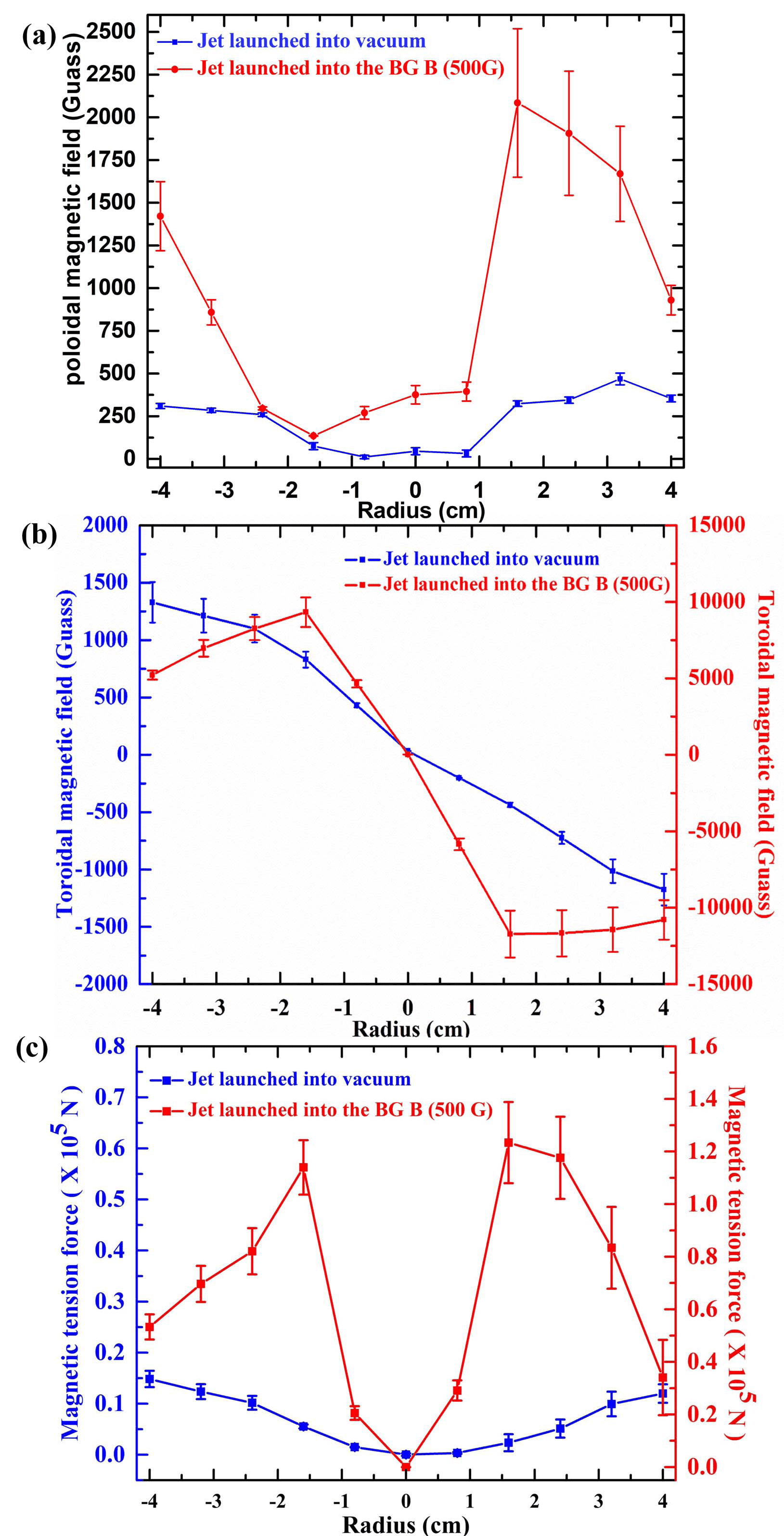}
\caption{(Color online).~Radial profile of (a) poloidal ($B_{z}$) magnetic field, (b) toroidal ($B_{\varphi}$) magnetic field and (c) the calculated magnetic tension force for background transverse magnetic field case and vacuum case respectively.~Comparing with vacuum case, both poloidal and toroidal magnetic field peak move toward the center (r = 0) due to the magnetic tension force, along with the magnitude intensified simultaneously}
\label{fig:Bcomp}
\end{figure}

The advected background magnetic field line is curved with the movement of the plasma jet.~Fig.~\ref{fig:Bcomp} shows both the poloidal and toroidal magnetic fields measured for this case compared with the vacuum case. The measured magnetic fields are intensified because the magnetic tension force of the background magnetic field works on the jet body, like an external pinch force, to compact the jet body, making it denser and more collimated. As a result, the magnetic flux density, $B$, and the current density, $J_{z}$, are increased leading to increased measurements for both the poloidal and toroidal magnetic fields.~As Fig.~\ref{fig:Bcomp} indicates, for the plasma jet propagating into vacuum case, in the B-dot probe measurement range, the locations of the peak toroidal magnetic field are at $r=\pm$4~cm, while for the jet launched into the background transverse magnetic field, the peaks are located at $r=\pm$1.5 cm.~This reduction in radius is convincing evidence for the pinch effect caused by the background magnetic tension force.~Furthermore, the radial profile of the magnetic tension force is calculated from measured magnetic field data, as shown in Fig.~\ref{fig:Bcomp}. The force peaks at at $r=\pm$1.5~cm, gradually reduces towards the center which is consistent with the spatial velocity shear profile.\cite{zhang2017emergent}

The effect of sheared flow on current-driven MHD instabilities has previously been investigated theoretically and demonstrated experimentally.\cite{shu1995sheared,shumlak2001evidence,shumlak2003sheared,golingo2005formation,shumlak2006plasma,shumlak2012sheared,shumlak2017increasing} The plasma jet equilibrium can be expressed by the MHD force balance equation:
\begin{equation}\label{eq:3}
\nabla{P}+\rho(\vec{\mathbf{V}}\cdot\nabla){\vec{\mathbf{V}}}=\vec{\mathbf{J}}\times\vec{\mathbf{B}}
\end{equation}
where $\nabla{P}$ is the plasma pressure, $\rho$ is the mass density, $\vec{\mathbf{V}}$ is the plasma velocity, $\vec{\mathbf{J}}$ is the current density and $\vec{\mathbf{B}}$ is the magnetic field.

For PBEX, assuming the pressure gradients are only radial, a constant initial axial magnetic field, $B_{z}$, is applied, and the plasma jet only propagates in the z-direction, Eq.~\ref{eq:3} can be simplified to:
\begin{equation}\label{eq:4}
\frac{B_{\varphi}}{\mu_{0}{r}} \frac{d(rB_{\varphi})}{dr}+\frac{dP}{dr}=0
\end{equation}

\begin{figure}[htb]
\includegraphics[width=0.5\textwidth]{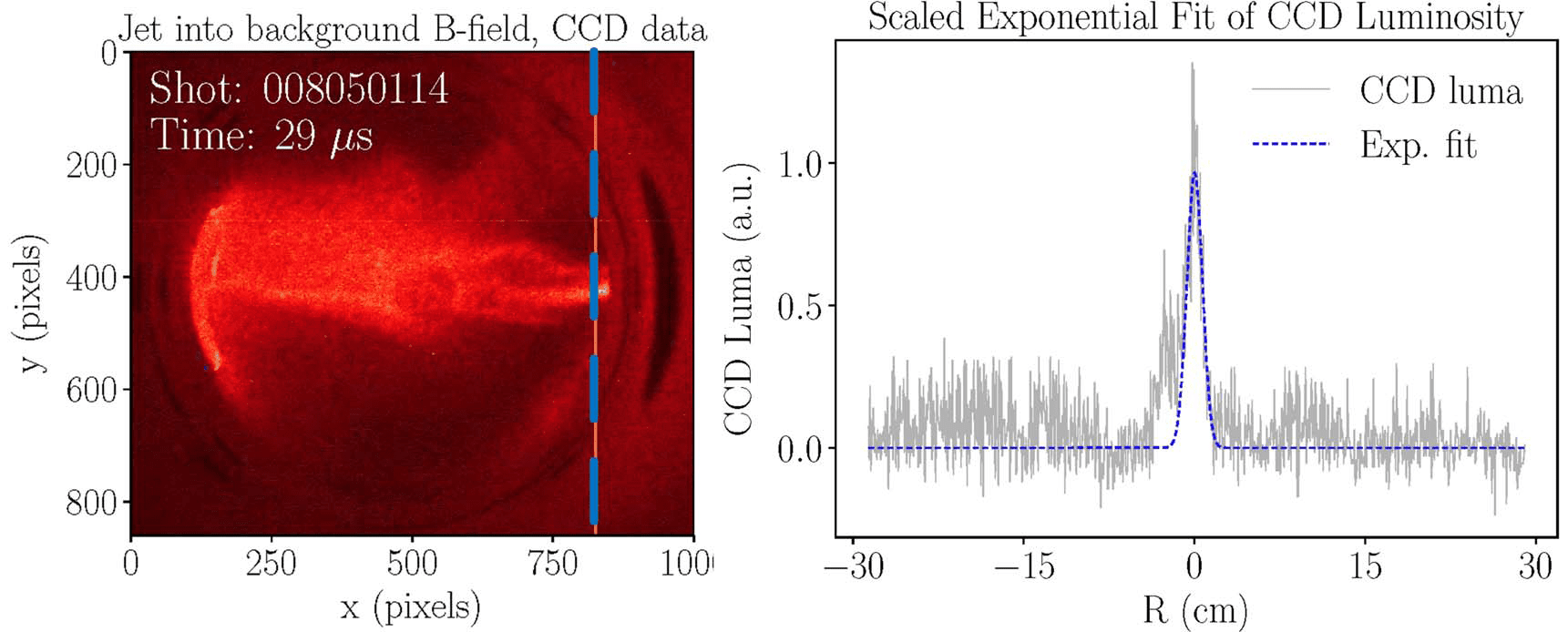}
\caption{(Color online).~plasma intensity radius-profile (spatial location is marked as blue color in the CCD camera picture on the left) with the fitting results under the Bennett equilibrium assumption}
\label{fig:pressurefit}
\end{figure}

Image data from CCD camera and magnetic field data from B-dot probe are utilized to analyze and examine the initial conditions of PBEX. There are two unknown terms, $P$ and $B_{\varphi}$, in Eq.~\ref{eq:4}. The plasma pressure term, $P$, is demonstrated from the CCD camera image data.~Assuming the plasma light intensity is positively related to the plasma density, for $T_{i}=T_{e}$, the plasma pressure is also positively correlated with the plasma density. As a result, the radial profile of the initial plasma pressure follows the plasma intensity from the image data as plotted in Fig.~\ref{fig:pressurefit}.

For constant electron drift velocity across the jet, with $T_{i}=T_{e}$, the Bennett equilibrium profile for the pressure, $P$, is of the form:
\begin{equation}\label{eq:5}
P_{0} = \frac{1}{\left (1+\frac{r^2}{a^2}\right )^2}
\end{equation}
where $r$ is the radius and $a$ is a parameter which sets the radius of the jet pressure profile.

Eq.~\ref{eq:5} satisfies the general radial force balance equation Eq.~\ref{eq:4}, and is employed to fit the plasma intensity profile shown in Fig.~\ref{fig:pressurefit}. The fitting result indicates that the Bennett equilibrium profile is a well fitting and reasonable assumption for the initial plasma pressure data. We can substitute Eq.~\ref{eq:5} with $a = 0.048$, which is determined by the fitting result shown in Fig.~\ref{fig:pressurefit}, into Eq.~\ref{eq:4}, to obtain the toroidal magnetic field, $B_{\varphi}$, profile as:
\begin{equation}\label{eq:6}
B_{0\varphi}=\sqrt{2}\frac{\frac{r}{a}}{1+(\frac{r}{a})^2}
\end{equation}

\begin{figure}[htb]
\includegraphics[width=0.5\textwidth]{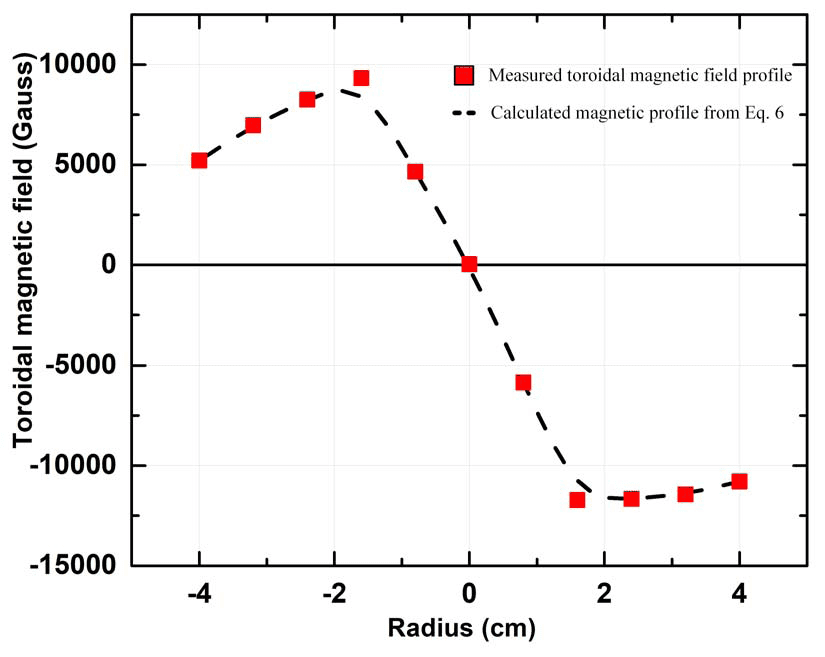}
\caption{(Color online).~Comparison between measured toroidal magnetic field radius-profile (red square) and the fitting results (dash line) calculated from Eq.~\ref{eq:6} under the Bennett equilibrium assumption}
\label{fig:magneticfit}
\end{figure}

The comparison between the calculated toroidal magnetic field profile from Eq.~\ref{eq:6} and the measured magnetic field data is plotted in Fig.~\ref{fig:magneticfit} where the experimental data show good consistency with the initial conditions of the model.

Based on the initial condition discussed above, the linearization of the MHD equations has shown that a sheared axial flow can stabilize the kink mode of the plasma jet and the required amount of the flow shear is given by:
\begin{equation}\label{eq:7}
\frac{dV_{Z}}{dr}\geq 0.1kV_{A}
\end{equation}
where $k$ is the axial wave number, and $V_{A}$ is the Alfv\'en velocity.~The experimental results from PBEX have been examined and the axial sheared flow's radial profile are consistent with Eq.~\ref{eq:7}.\cite{zhang2017emergent}

\subsection{Plasma bubble injection into background transverse magnetic field}
\subsubsection{Identification of spheromak formation}
\begin{figure}
\includegraphics[width=0.5\textwidth]{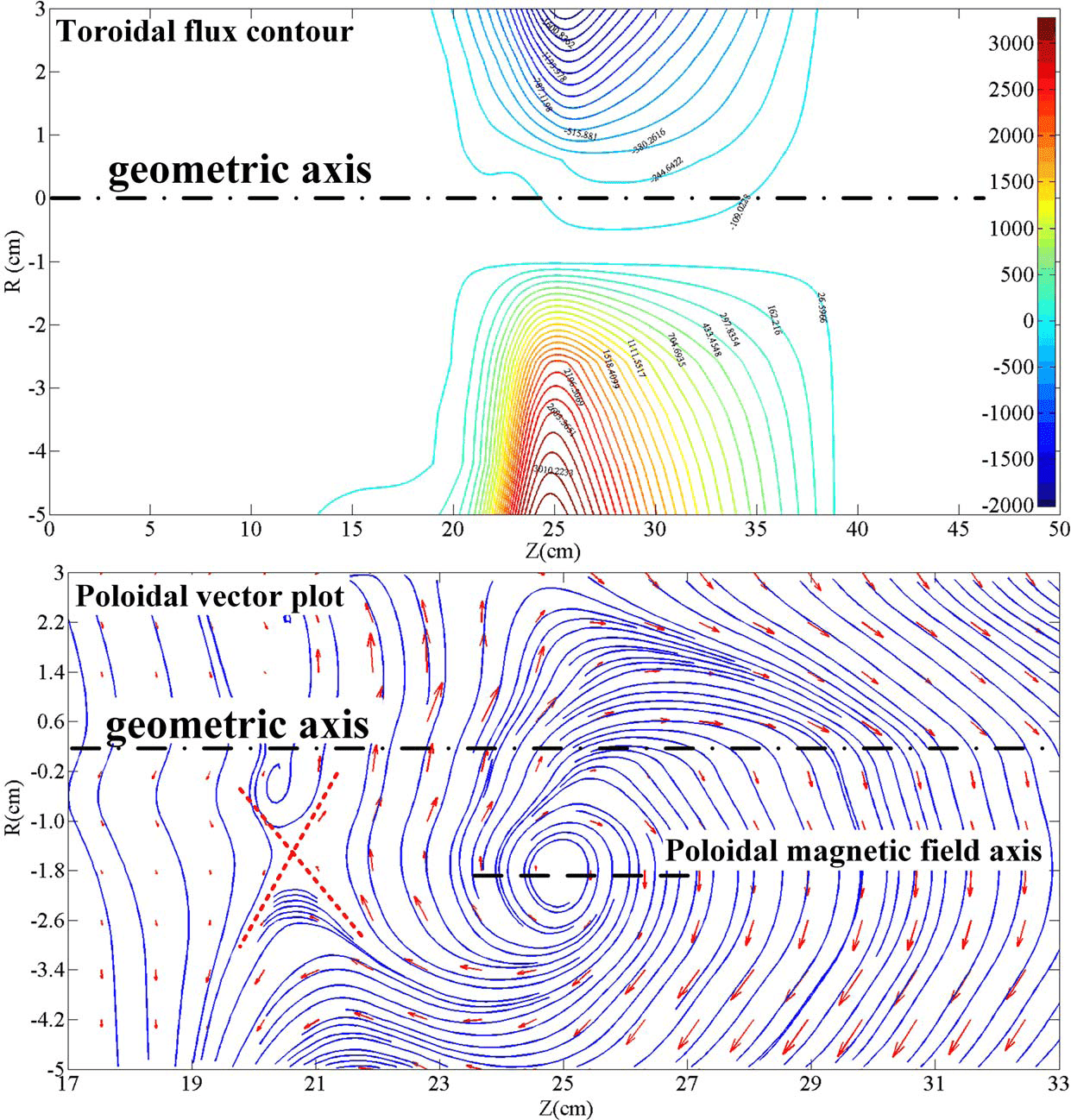}
\caption{(Color online).~Typical spheromak self-closed toroidal contour plot and poloidal vector plot with magnetic reconnection X point detected.}
\label{fig:bubblemagneticfield}
\end{figure}

Before launching a plasma bubble into a background transverse magnetic field, a self-organized spheromak formation is firstly identified with prior work.\cite{yamada1981quasistatic,jarboe1994review} A typical spheromak image sequence is shown in Fig.~\ref{fig:formations} (b) where a bright leading surface propagates across the whole main-chamber cross-section. The measured toroidal contour plot and poloidal vector plot are shown in the Fig.~\ref{fig:bubblemagneticfield}. The characteristic self-closed magnetic field configurations are detected for both toroidal and poloidal magnetic field. We note here that an "X" point is detected in the poloidal vector plot (marked in Fig.~\ref{fig:bubblemagneticfield}) after a bubble detaches the gun nuzzle, indicating the occurrence of magnetic reconnection during the detachment.

\subsubsection{Observation of finger-like structures at the bubble edge}
\begin{figure}
\includegraphics[width=0.5\textwidth]{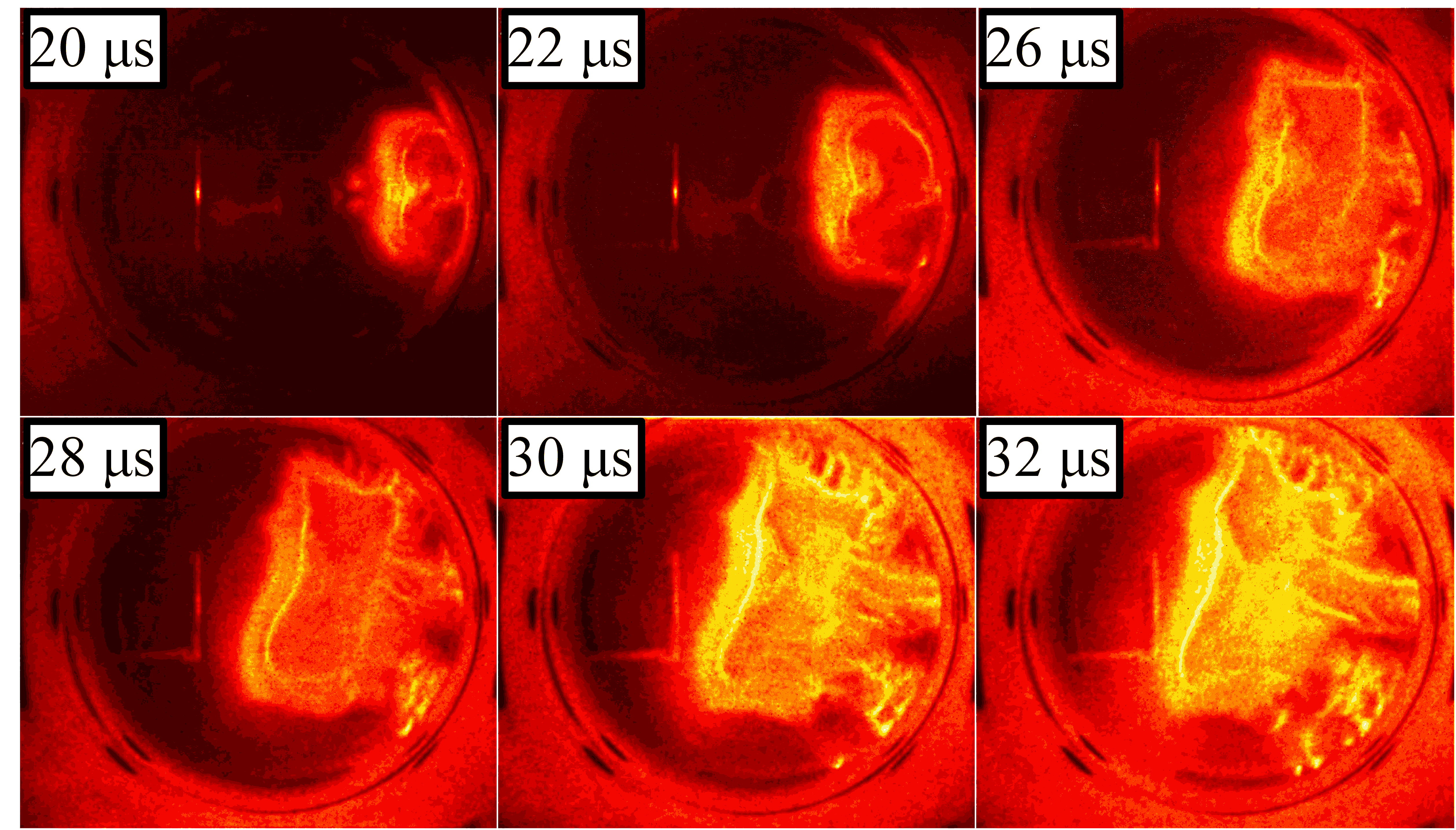}
\caption{(Color online).~Typical plasma bubble propagates into background transverse magnetic field images. The background magnetic field direction is pointing into the page. The finger-like-structure appears on the top side where there is a strong magnetic field null between the spheromak and background field~(Pulse No.\ 034050914)}
\label{fig:bgbubble}
\end{figure}

A series of plasma evolution images is shown in Fig.~\ref{fig:bgbubble} for the case of a coaxial-gun-formed bubble launched into a background transverse magnetic field (500 Gauss). As the image data shows, at the first stage, similar to the vacuum case, the gun-generated plasma follows the bias magnetic field lines, and then a typically symmetric spheromak shape is formed around $t = 20$~$\mu$s. Then at $t = 22$~$\mu$s, the spheromak starts to propagate into the chamber with a transverse 500-G background magnetic field. The plasma does not travel strictly along the z-axis as it does in the vacuum case in Fig.~\ref{fig:formations} (b). Instead, the bubble deviates from the z-axis and moves upwards. The typical spheromak closed-field configuration does not hold anymore. Furthermore, during the time period from $t = 26$~$\mu$s to $t = 32$~$\mu$s, on the upper side, finger-like structures develop at the interface between the plasma and the background magnetic field. The structures are up-down asymmetric and reach the vacuum chamber boundary during dynamic evolution.

\begin{figure}
\includegraphics[width=0.35\textwidth]{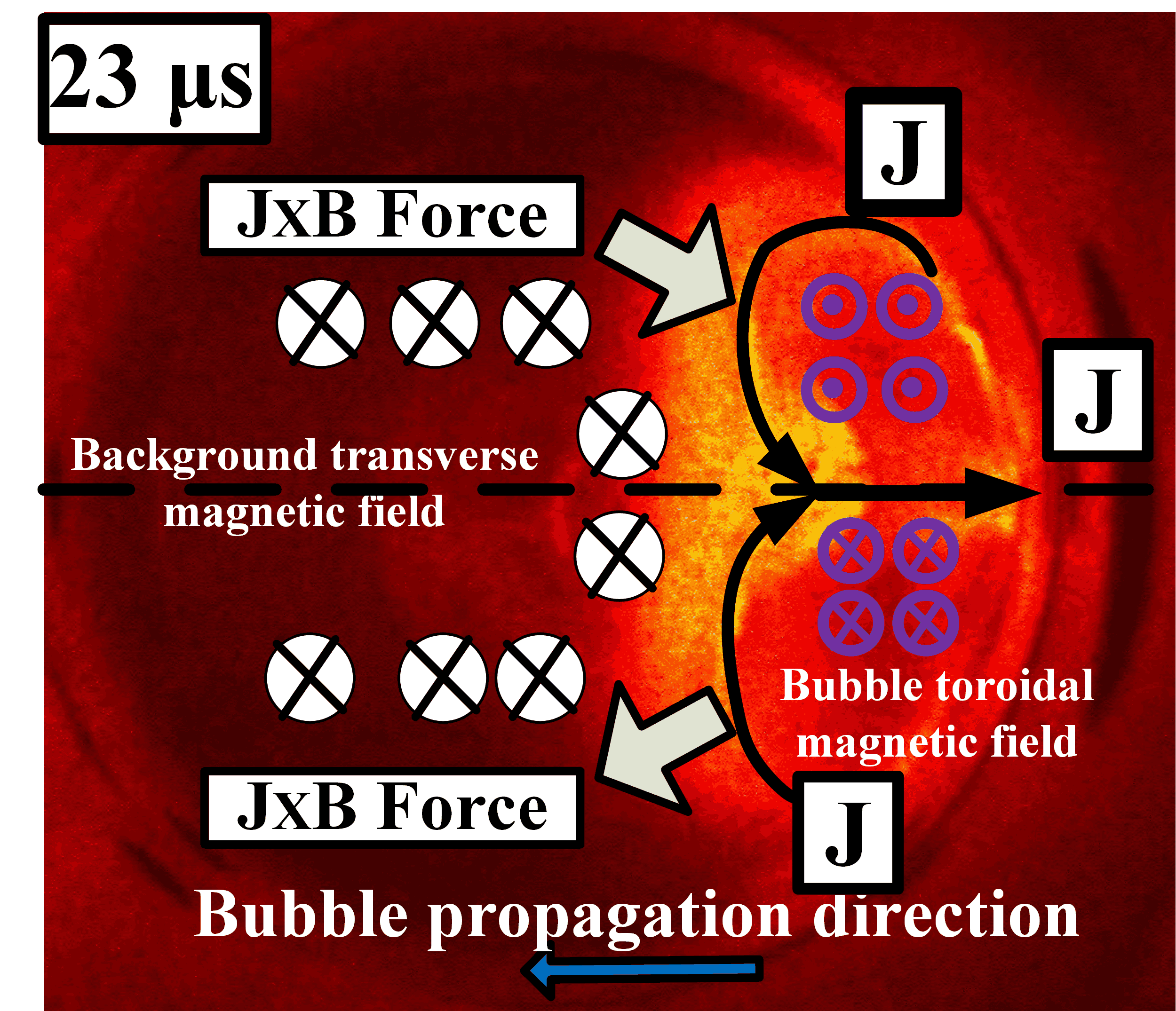}
\caption{(Color online).~The background magnetic field and the spheromak toroidal magnetic field configurations. There is a stronger magnetic field null on the top-side between the background magnetic field and spheromak magnetic field itself}
\label{fig:physcisMRT}
\end{figure}

A simple physical picture is presented in Fig.~\ref{fig:physcisMRT} with the configurations of both $B_{background}$ and $B_{toroidal}$. Based on the 2D image data, at the interface between the plasma and the background magnetic fields, for the upper part, these two magnetic field regions are antiparallel to each other and vice versa for the lower part. 

Based on the observed finger-like structures, mRT is a possible explanation.~It is well known that the formation and evolution of the mRT instability in a magnetized plasma heavily depends on the geometry of the magnetic field relative to the interface (between the plasma and the background magnetic field for PBEX).\cite{treumann1997advanced,dursi2008draping,moser2012magnetic} The growth rate of the two-dimensional (2D) mRT instability is expressed below:\cite{harris1962rayleigh}
\begin{equation}\label{eq:8}
\gamma^{2}=\vec{\alpha}\cdot \vec{k}-\frac{(\vec{k}\cdot\vec{B})^{2}}{\mu _{0}{\rho _{plasma}}}
\end{equation}
where $\vec{B}$ is the unperturbed magnetic field vector, $\vec{\alpha}$ is the acceleration, $\vec{k}$ is the perturbation wave-vector, and $\rho _{plasma}$ is the plasma mass density.~In the case of a magnetic field perpendicular to the surface, the mRT instability will reduce to the unmagnetized fluid case.~The magnetic field has no effect for large wavelength disturbances. 

In the case of the magnetic field parallel to the interface (PBEX case), Eq.~\ref{eq:8} indicates that the term, $\vec{k}\cdot\vec{B}$, contributes to the stabilization, and suppresses the instability growth rate as discussed above. For the upper part of the plasma, the $B_{background}$ and $B_{toroidal}$ are antiparallel and there is a strong magnetic field null at the interface as shown in Fig.~\ref{fig:physcisMRT}. Consequently, the stabilization term, $\vec{k}\cdot\vec{B}$, is reduced and the mRT instability is more likely to happen at the upper side rather than the lower side.

\begin{figure}
\includegraphics[width=0.35\textwidth]{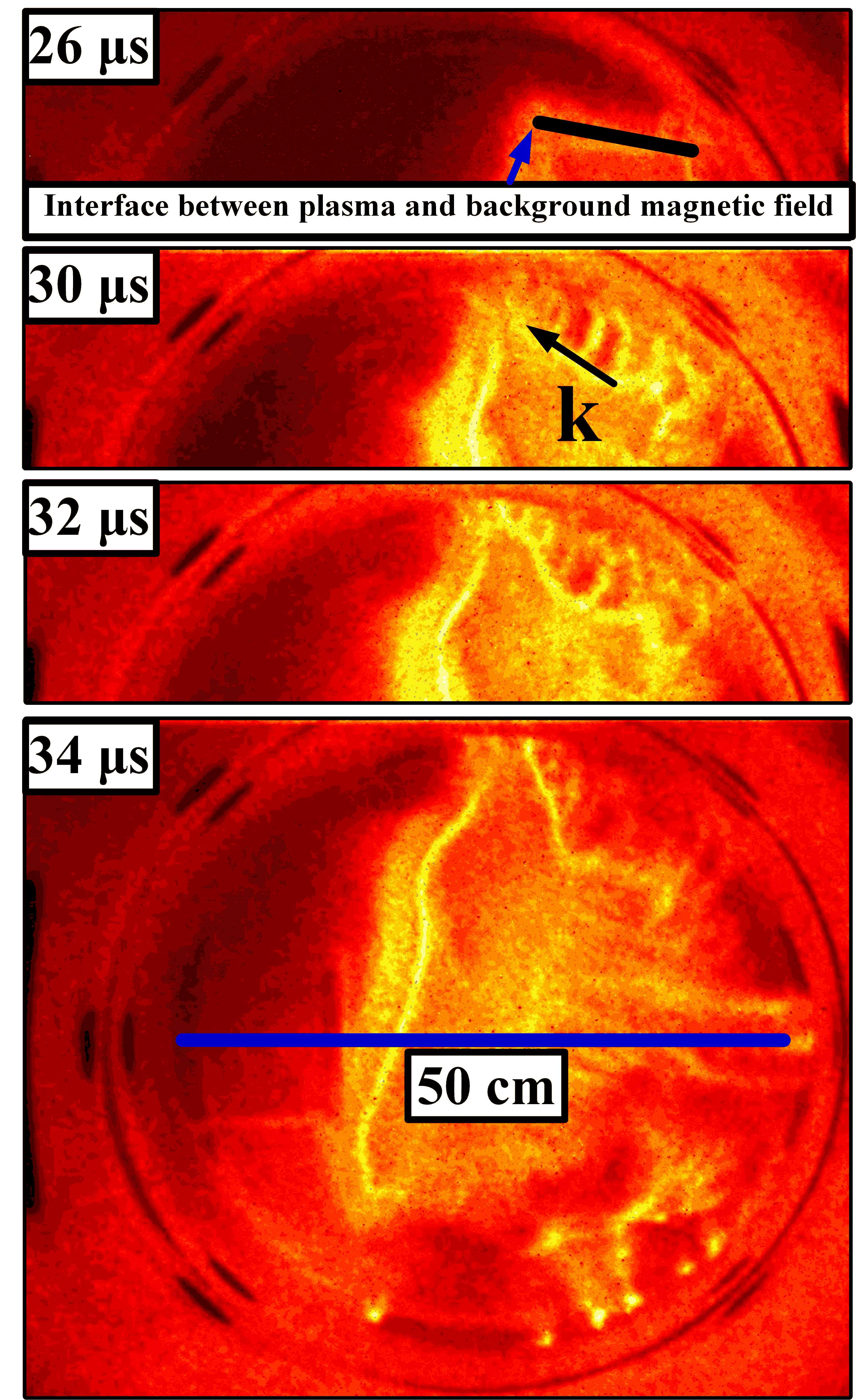}
\caption{(Color online).~The magnetic Rayleigh-Taylor instability growth rate calculation based on the image data (Pulse No.\ 034050914).}
\label{fig:MRTcal}
\end{figure}

The mRT instability growth rate can be estimated directly from the image data.~Based on the image data shown in Fig.~\ref{fig:MRTcal}, the amplitude time evolution of the finger-like structure  gives an observed growth rate $\gamma _{mRT}\approx 1.43\times 10^{6}~s^{-1}$. Also from the images, the measured transverse acceleration of the filament is: $\alpha \approx 1.78\times 10^{10}$~$m/s^{2}$. The axial wavenumber, $\vec{k}$, is 503 $m^{-1}$. Then the calculated growth rate is: $\gamma_{mRT-c}=\sqrt{\vec{\alpha}\cdot \vec{k}}=3.0\times 10^{6}$ $s^{-1}$. Both calculations are in good agreement with each other. The growth rate agreement, along with the observed location, and spatial periodicity of the finger structures are all consistent with the mRT instability.\cite{chandrasekhar1961hydrodynamic}

\begin{figure}
\includegraphics[width=0.5\textwidth]{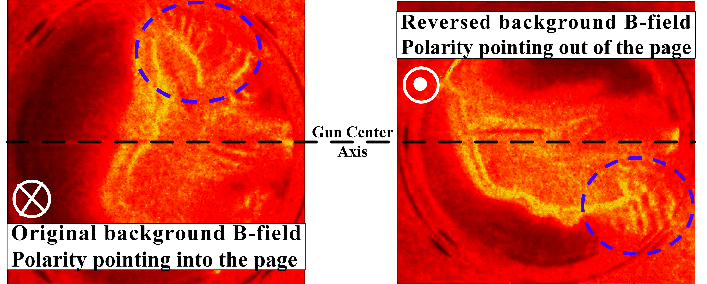}
\caption{(Color online).~The comparison of CCD camera image data for spheromak launched into the background magnetic field with original polarity (pulse No.\ 034050914) and with reversed polarity (Pulse No.\ 008102816).}
\label{fig:MRTreverse}
\end{figure}

To further examine this suggested explanation, the background magnetic field direction is reversed, pointing anti-parallel to the source (out of the paper) instead of the original source-directed background field (into the paper). For this case, the stabilization term, $\vec{k}\cdot\vec{B}$, is reduced and the mRT instability is more likely to happen at the lower side rather than the supper side. The CCD camera images for this reversed-field case is shown in Fig.~\ref{fig:MRTreverse}, compared with the original background magnetic field setting.

Furthermore, when the closed-flux bubble expands into the background transverse magnetic field, the $B_{background}$ and $B_{toroidal}$ are antiparallel as shown in Fig.~\ref{fig:physcisMRT}. This could lead to magnetic reconnection and a current-density layer at the interface.~Thus, what we observe may also be related to current-driven, nonlinear reconnecting edge localized modes.\cite{ebrahimi2017nonlinear}~However, further measurements are needed to explore this possibility, and to distinguish between the pressure-driven (mRT) and current-driven mechanisms.

\section{Plasma-jet and bubble injection into background plasma}
\begin{figure}
\includegraphics[width=0.5\textwidth]{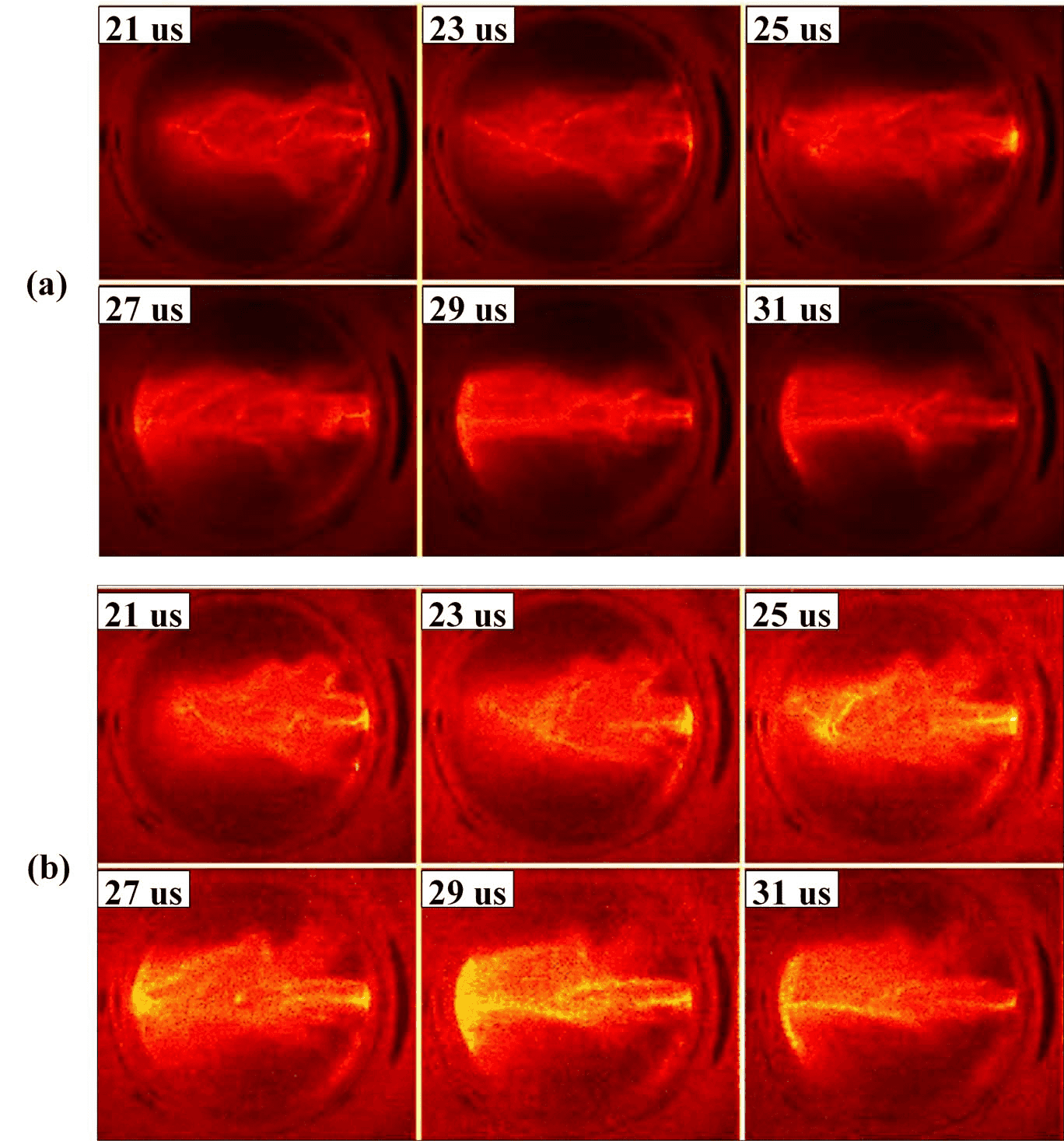}
\caption{(Color online).~(a) CCD camera image data for plasma jet launched into the background magnetized plasma (pulse No.\ 007050114).~(b) CCD camera image data for plasma jet launched into the background magnetized field (Pulse No.\ 008050114).}
\label{fig:jetcomparision}
\end{figure}

\begin{figure}
\includegraphics[width=0.5\textwidth]{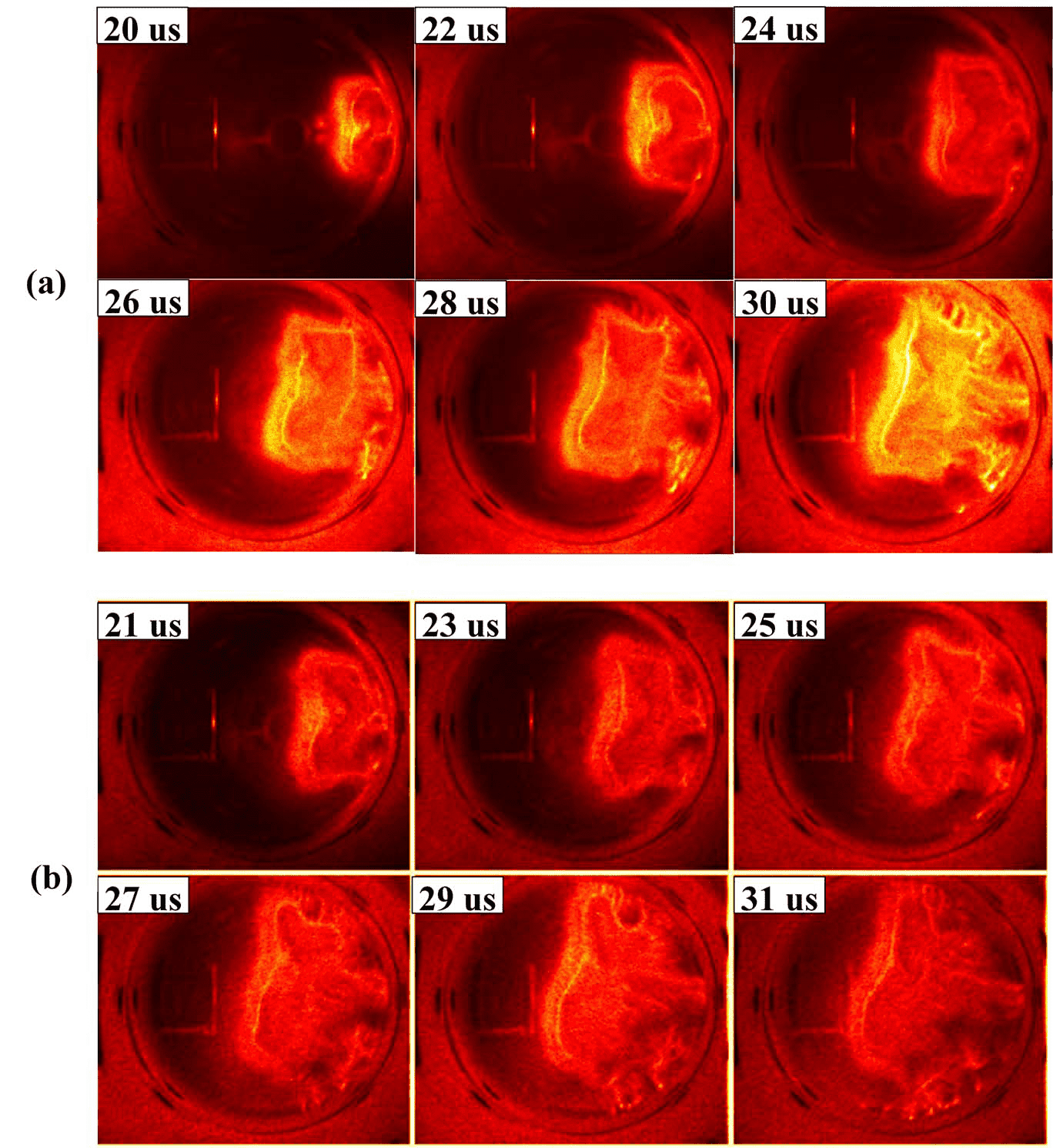}
\caption{(Color online).~(a) CCD camera image data for plasma bubble launched into the background magnetized field (pulse No.\ 034050914).~(b) CCD camera image data for plasma jet launched into the background magnetized plasma (Pulse No.\ 035050914).}
\label{fig:bubblecomparision}
\end{figure}

\begin{figure}
\includegraphics[width=0.5\textwidth]{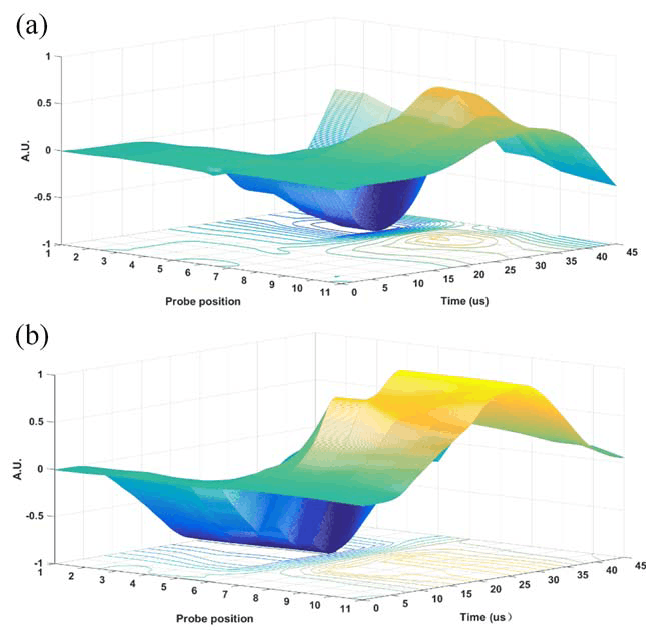}
\caption{(Color online).~(a) Toroidal magnetic field plot for plasma jet launched into the background magnetized field (pulse No.\ 007050114).~(b) Toroidal magnetic field plot for plasma jet launched into the background magnetized plasma (Pulse No.\ 008050114).}
\label{fig:jetBcomparision}
\end{figure}

As discussed in Sec.~I, for low-$\beta$ background magnetized plasma case, the background magnetic field plays the most important role during the interaction process. However, we did launch gun-formed plasmas into the background plasma to verify this assumption. The initial CCD camera images show very similar plasma dynamics, as shown in Fig.~\ref{fig:jetcomparision} and \ref{fig:bubblecomparision} for the jet and bubble cases, respectively. 
Furthermore, magnetic probe measurements also show the similar global magnetic field configurations, as shown in Fig.~\ref{fig:jetBcomparision} for the plasma-jet case (as an example).~The experimental data indicates that the background magnetic field provides the dominant mechanism for the observations reported in this paper.

\section{Summary}
In this paper, gun-formed plasma jets and bubbles (spheromaks) propagating into background transverse magnetic field or background plasma are discussed. Both the CCD camera and B-dot probe data show that the primary observations reported here are similar in both cases, indicating that the background field dominates the interaction. As a result, this paper has focused on presenting data from the background-field-only case.

Plasma jets propagating into a background transverse magnetic field are observed to be more stable than into vacuum. The calculated magnetic Reynolds number indicates that the background magnetic field is advected with the plasma flow.~The bending magnetic field lines create a magnetic tension force that acts on the jet body, pinching the jet body column, and causing a significant shear in the axial flow.~The measured poloidal and toroidal magnetic field radial profile, along with the calculated magnetic tension force profile, confirms the pinch effects. However, evaluation of the Kruskal-Shafranov criterion based on the measured safety factor indicates that the jet should be kink unstable; linearization of MHD equations shows that if the axial sheared flow is greater than the criterion, $0.1kV_{A}$, it can suppress, or reduce the $n=1$ kink instability, making the jet more stable. The measured axial sheared flow data are consistent with the theoretical analysis.\cite{zhang2017emergent}

Significant differences are also found in the case of spheromak plasmas propagating into background magnetic field as compared with vacuum. In the former case, non-uniform (i.e., up-down asymmetry) expansion, and finger-like structures are observed at the upper side in CCD-camera images. A simple physical picture is proposed to describe how the background magnetic field affects the spheromak. The observed non-uniform plasma expansion (up-down non-uniformity) is due to different $\vec{J}\times \vec{B}$ forces.~The measured poloidal and toroidal magnetic field data show that the plasma does not hold the typical self-closed magnetic configuration under the influence of a background magnetic field. On the upper side, the opposite directions of the background and the toroidal magnetic field reduce the stabilization term $\vec{k}\cdot\vec{B}$ which makes it easier for the mRT to happen at this side. The growth rate is calculated from the image data, and the results indicate that the mRT growth rate is in the order of $10^{6}~s^{-1}$ for PBEX. Another possible explanation is that a current-density layer forming at the bubble/background interface gives rise to growth of current-driven unstable filaments,\cite{ebrahimi2017nonlinear} but further measurements would be needed to verify this possibility.

The authors acknowledge Dr.\ Glen Wurden for loaning the CCD camera, Dr.\ Kevin Yates for assistance in camera setup and operation, and Dr.\ Fatima Ebrahimi for pointing out the possibility that the observed finger-like structures could potentially be current-driven, unstable filaments akin to edge-localized modes. 

This material is based upon work supported by the National Science Foundation under Grant No. AST-0613577 and the Army Research Office under award No. W911NF1510480.


\begin{thebibliography}{}

\bibitem{honda2002self}
Mitsuru Honda and Yasuko~S Honda.
"Self-collimation and magnetic field generation of astrophysical jets,"
\href{http://dx.doi.org/10.1086/340455} {Astrophys. J. Lett.}, 569(1):L39, 2002.

\bibitem{pelletier1992hydromagnetic}
Guy Pelletier and Ralph~E Pudritz.
"Hydromagnetic disk winds in young stellar objects and active galactic nuclei,"
\href{http://dx.doi.org/10.1086/171565} {Astrophys. J.}, 394:117--138, 1992.

\bibitem{clausen2011signatures}
E. Clausen-Brown, M. Lyutikov and P. Kharb. 
"Signatures of large-scale magnetic fields in active galactic nuclei jets: transverse asymmetries,"
\href{http://dx.doi.org/10.1111/j.1365-2966.2011.18757.x} {Mon. Not. R. Astron. Soc.}, 415(3), pp.2081-2092, 2011

\bibitem{schatten1969model}
Kenneth~H Schatten, John~M Wilcox, and Norman~F Ness.
"A model of interplanetary and coronal magnetic fields,"
\href{http://dx.doi.org/10.1007/BF00146478} {Sol. Phys.}, 6(3):442--455, 1969.

\bibitem{jokipii1990polar}
RJ~Jokipii and J~Kota.
"The polar heliospheric magnetic field,"
\href{http://dx.doi.org/10.1029/GL016i001p00001}{Geophys. Res. Lett.}, volume~16, page 1-4, 1989.

\bibitem{zmuda1970characteristics}
AJ~Zmuda, JC~Armstrong, and FT~Heuring.
"Characteristics of transverse magnetic disturbances observed at 1100 kilometers in the auroral oval,"
\href{https://doi.org/10.1029/JA075i025p04757} {J. Geophys. Res.}, 75(25):4757--4762, 1970.

\bibitem{lyons2013quantitative}
Larry~R Lyons and Donald~J Williams.
"Quantitative aspects of magnetospheric physics (Volume~23),"
\href{http://www.springer.com/us/book/9789027716637} {Springer Science \& Business Media}, 2013.

\bibitem{leonard1986trapping}
A. W. Leonard, R. N. Dexter and J.C.Sprott. 
"Trapping of gun-injected plasma by a tokamak,"
\href{https://doi.org/10.1103/PhysRevLett.57.333} {Phys. Rev. Lett.}, 57(3), p.333, 1986.

\bibitem{brown1990current}
M. R. Brown and P. M. Bellan.
"Current drive by spheromak injection into a tokamak,"
\href{https://doi.org/10.1103/PhysRevLett.64.2144} {Phys. Rev. Lett.}, 64(18):2144, 1990.

\bibitem{brown1990spheromak}
M. R. Brown and P. M. Bellan.
"Spheromak injection into a tokamak,"
\href{http://dx.doi.org/10.1063/1.859546} {Phys. Fluids B}, 2(6):1306--1310, 1990.

\bibitem{raman1994experimental}
R.~Raman, F.~Martin, B.~Quirion, M.~St-Onge, J.~L.~Lachambre, D.~Michaud, B.~Sawatzky, J.~Thomas, A.~Hirose, D.~Hwang and N.~Richard. 
"Experimental demonstration of nondisruptive, central fueling of a tokamak by compact toroid injection,"
\href{https://doi.org/10.1103/PhysRevLett.73.3101} {Phys. Rev. Lett.}, 73(23), p.3101, 1994.

\bibitem{loewenhardt1995performance}
P. K. Loewenhardt, M. R. Brown, J. Yee and P. M. Bellan.
"Performance characterization of the Caltech compact torus injector,"
\href{https://doi.org/10.1063/1.1146044}{Rev. Sci. Instrum.}, 66(2), pp.1050-1055, 1995.

\bibitem{raman1997experimental}
R. Raman, F. Martin, E. Haddad, M. St-Onge, G. Abel, C. Cote, N. Richard, N. Blanchard, H. H. Mai, B. Quirion, et~al.
"Experimental demonstration of tokamak fuelling by compact toroid injection,"
\href{http://dx.doi.org/10.1088/0029-5515/37/7/I05}{Nucl. Fusion}, 37(7):967, 1997.

\bibitem{mclean1998design}
H. S. McLean, D. Q. Hwang, R. D. Horton, R. W. Evans, S. D. Terry, J. C. Thomas and R. Raman.
"Design and operation of a passively switched repetitive compact toroid plasma accelerator,"
\href{https://doi.org/10.13182/FST98-A31}{Fusion Technol.}, 33(3), pp.252-272, 1998.

\bibitem{olynyk2008development}
G. Olynyk and J. Morelli.
"Development of a compact toroid fuelling system for ITER,"
\href{https://doi.org/10.1088/0029-5515/48/9/095001}{Nucl. Fusion}, 48(9):095001, 2008.

\bibitem{ebrahimi2016dynamo}
F. Ebrahimi. 
"Dynamo-driven plasmoid formation from a current-sheet instability,"
\href{https://doi.org/10.1063/1.4972218}{Phys. Plasmas}, 23(12), p.120705, 2016.

\bibitem{ebrahimi2017nonlinear}
F. Ebrahimi.
"Nonlinear reconnecting edge localized modes in current-carrying plasmas,"
\href{https://doi.org/10.1063/1.4983631}{Phys. Plasmas}, 24(5), p.056119, 2017

\bibitem{tuck1959plasma}
J.~L.~Tuck. 
"Plasma jet piercing of magnetic fields and entropy trapping into a conservative system,"
\href{https://doi.org/10.1103/PhysRevLett.3.313} {Phys. Rev. Lett.}, 3(7), p.313, 1959.

\bibitem{schmidt1960plasma}
George Schmidt.
"Plasma motion across magnetic fields,"
\href{http://dx.doi.org/10.1063/1.1706163} {Phys. Fluids}, 3(6):961--965, 1960.

\bibitem{baker1965experimental}
D. A.~Baker and J. E.~Hammel.
"Experimental studies of the penetration of a plasma stream into a transverse magnetic field,"
\href{http://dx.doi.org/10.1063/1.1761288} {Phys. Fluids}, 8(4):713--722, 1965.

\bibitem{parks1988refueling}
P. B. Parks. 
"Refueling tokamaks by injection of compact toroids,"
\href{https://doi.org/10.1103/PhysRevLett.61.1364} {Phys. Rev. Lett.}, 61(12), p.1364, 1988.

\bibitem{hsu2002laboratory}
S. C. Hsu and P.~M.~Bellan.
"A laboratory plasma experiment for studying magnetic dynamics of accretion discs and jets,"
\href{https://doi.org/10.1046/j.1365-8711.2002.05422.x} {Mon. Not. R. Astron. Soc.}, 334(2):257--261, 2002.

\bibitem{hsu2003experimental}
S. C. Hsu and P.~M.~Bellan.
"Experimental identification of the kink instability as a poloidal flux amplification mechanism for coaxial gun spheromak formation,"
\href{https://doi.org/10.1103/PhysRevLett.90.215002} {Phys. Rev. Lett.}, 90(21):215002, 2003.

\bibitem{hsu2005jets}
S. C. Hsu and P.~M.~Bellan.
"On the jets, kinks, and spheromaks formed by a planar magnetized coaxial gun,"
\href{http://dx.doi.org/10.1063/1.1850921} {Phys. Plasmas}, 12(3):032103, 2005.

\bibitem{bellan2005simulating}
Paul~M Bellan, Setthivoine You, and Scott~C Hsu.
"Simulating astrophysical jets in laboratory experiments,"
\href{http://dx.doi.org/10.1007/s10509-005-3933-1} {Astrophys. Space Sci}, 298(1):203--209, 2005.

\bibitem{yun2007large}
Gunsu~S Yun, Setthivoine You, and Paul~M Bellan.
"Large density amplification measured on jets ejected from a magnetized plasma gun,"
\href{http://dx.doi.org/10.1088/0029-5515/47/3/003} {Nucl. Fusion}, 47(3):181, 2007.

\bibitem{lebedev2002laboratory}
SV~Lebedev, JP~Chittenden, FN~Beg, SN~Bland, A~Ciardi, D~Ampleford, S~Hughes, MG~Haines, A~Frank, EG~Blackman, et~al.
"Laboratory astrophysics and collimated stellar outflows: The production of radiatively cooled hypersonic plasma jets,"
\href{https://doi.org/10.1086/324183} {Astrophys. J.}, 564(1):113, 2002.

\bibitem{lebedev2004jet}
SV~Lebedev, D~Ampleford, A~Ciardi, SN~Bland, JP~Chittenden, MG~Haines, A~Frank, EG~Blackman, and A~Cunningham.
"Jet deflection via crosswinds: Laboratory astrophysical studies,"
\href{https://doi.org/10.1086/423730} {Astrophys. J.}, 616(2):988, 2004.

\bibitem{lebedev2005magnetic}
SV~Lebedev, A~Ciardi, DJ~Ampleford, SN~Bland, SC~Bott, JP~Chittenden, GN~Hall, J~Rapley, CA~Jennings, A~Frank, et~al.
"Magnetic tower outflows from a radial wire array z-pinch,"
\href{http://dx.doi.org/10.1111/j.1365-2966.2005.09132.x} {Mon. Not. R. Astron. Soc.}, 361(1):97--108, 2005.

\bibitem{ciardi2007evolution}
A~Ciardi, SV~Lebedev, A~Frank, EG~Blackman, JP~Chittenden, CJ~Jennings, DJ~Ampleford, SN~Bland, SC~Bott, J~Rapley, et~al.
"The evolution of magnetic tower jets in the laboratory,"
\href{http://dx.doi.org/10.1063/1.2436479} {Phys. Plasmas}, 14(5):056501, 2007.

\bibitem{mostovych1989laser}
AN~Mostovych, BH~Ripin, and JA~Stamper.
"Laser produced plasma jets: Collimation and instability in strong transverse magnetic fields,"
\href{https://doi.org/10.1103/PhysRevLett.62.2837} {Phys. Rev. Lett.}, 62(24):2837, 1989.

\bibitem{ripin1990laboratory}
BH~Ripin, CK~Manka, TA~Peyser, EA~McLean, JA~Stamper, AN~Mostovych, J~Grun, K~Kearney, JR~Crawford, and JD~Huba.
"Laboratory laser-produced astrophysical-like plasmas,"
\href{https://doi.org/10.1017/S026303460000793X} {Laser Part Beams}, 8(1-2):183--190, 1990.

\bibitem{harilal2004confinement}
SS~Harilal, MS~Tillack, B~O'Shay, CV~Bindhu, and F~Najmabadi.
"Confinement and dynamics of laser-produced plasma expanding across a transverse magnetic field,"
\href{https://doi.org/10.1103/PhysRevE.69.026413} {Phys Rev E Stat Nonlin Soft Matter Phys}, 69(2):026413, 2004.

\bibitem{matsumoto2016characterization}
T. Matsumoto, T. Roche, I. Allfrey, J. Sekiguchi, T. Asai, H. Gota, M. Cordero, E. Garate, J. Kinley, T. Valentine, et~al.
"Characterization of compact-toroid injection during formation, translation, and field penetration,"
\href{https://doi.org/10.1063/1.4959571} {Rev. Sci. Instrum.}, 87(11), p.11D406, 2016.

\bibitem{liu2008ideal}
Wei Liu, Scott~C Hsu, Hui Li, Shengtai Li, and Alan~G Lynn.
"Ideal magnetohydrodynamic simulation of magnetic bubble expansion as a model for extragalactic radio lobes,"
\href{http://dx.doi.org/10.1063/1.2948347} {Phys. Plasmas}, 15(7):072905, 2008.

\bibitem{liu2009ideal}
Wei Liu, Scott~C Hsu, and Hui Li.
"Ideal magnetohydrodynamic simulations of low beta compact toroid injection into a hot strongly magnetized plasma,"
\href{http://dx.doi.org/10.1088/0029-5515/49/9/095008} {Nucl. Fusion}, 49(9):095008, 2009.

\bibitem{liu2011ideal}
Wei Liu and Scott~C Hsu.
"Ideal magnetohydrodynamic simulations of unmagnetized dense plasma jet injection into a hot strongly magnetized plasma,"
\href{https://doi.org/10.1088/0029-5515/51/7/073026} {Nucl. Fusion}, 51(7):073026, 2011.

\bibitem{zhang2017emergent}
Y.~Zhang, M.~Gilmore, S.~C.~Hsu, D.~M.~Fisher, and A.~G.~Lynn.
"Emergent kink stability of a magnetized plasma jet injected into a transverse background magnetic field,"
\href{https://doi.org/10.1063/1.5010188} {Phys. Plasmas}, 24, 110702, 2017.

\bibitem{shu1995sheared}
U.~Shumlak and C.~W.~Hartman.
"Sheared flow stabilization of the m= 1 kink mode in z pinches,"
\href{https://doi.org/10.1103/PhysRevLett.75.3285} {Phys. Rev. Lett.}, 75(18):3285, 1995.

\bibitem{desjardins2016dynamics}
TR~Desjardins and M~Gilmore.
"Dynamics of flows, fluctuations, and global instability under electrode biasing in a linear plasma device,"
\href{http://dx.doi.org/10.1063/1.4948282} {Phys. Plasmas}, 23(5):055710, 2016.

\bibitem{kelly2016ari}
RF~Kelly, KD~Meaney, M~Gilmore, TR~Desjardins, and Y~Zhang.
"Ar\uppercase\expandafter{\romannumeral1}/Ar\uppercase\expandafter{\romannumeral2} laser induced fluorescence system for measurement of neutral and ion dynamics in a large scale helicon plasma,"
\href{http://dx.doi.org/10.1063/1.4959157} {Rev. Sci. Instrum.}, 87(11):11E560, 2016.

\bibitem{yee2000taylor}
J. Yee and P. M. Bellan.
"Taylor relaxation and $\lambda$ decay of unbounded, freely-expanding spheromaks,"
\href{https://doi.org/10.1063/1.1287137} {Phys. Plasmas}, 7(9), pp.3625-3640, 2000

\bibitem{moser2012magnetic}
Moser, A.L. and Bellan, P.M.
"Magnetic reconnection from a multiscale instability cascade,"
\href{https://doi.org/10.1038/nature10827} {Nature}, 482(7385), pp.379-381, 2012.

\bibitem{lynn2009helcat}
Alan~G Lynn, Mark Gilmore, Christopher Watts, Janis Herrea, Ralph Kelly, Steve Will, Shuangwei Xie, Lincan Yan, and Yue Zhang.
"The helcat dual-source plasma device,"
\href{http://dx.doi.org/10.1063/1.3233938} {Rev. Sci. Instrum.}, 80(10):103501, 2009.

\bibitem{gilmore2015helcat}
M~Gilmore, AG~Lynn, TR~Desjardins, Y~Zhang, C~Watts, SC~Hsu, S~Betts, R~Kelly, and E~Schamiloglu.
"The helcat basic plasma science device,"
\href{https://doi.org/10.1017/S0022377814000919} {J. Plasma Phys.}, 81(01):345810104, 2015.

\bibitem{zhang2009design}
Y.~Zhang, A.~G.~Lynn, S.~C.~Hsu, M.~Gilmore, and C.~Watts.
"Design of a compact coaxial magnetized plasma gun for magnetic bubble expansion experiments,"
\href{http://dx.doi.org/10.1109/PPC.2009.5386311} {PPC'09. IEEE}, pp.233--238. IEEE, 2009.

\bibitem{thomas1993simple}
J. C. Thomas, D. Q. Hwang, R. D. Horton, J. H. Rogers and R. Raman. 
"A simple fast pulse gas valve using a dynamic pressure differential as the primary closing mechanism,"
\href{https://doi.org/10.1063/1.1144053} {Rev. Sci. Instrum.}, 64(6), pp.1410-1413, 1993.

\bibitem{UHSi1224}
See \href{http://www.invisiblevision.com/products/ultra-high-speed-framing}{http://www.invisiblevision.com/products/ultra-high-speed-framing}
for UHSi 12/24 Key Features.

\bibitem{chapman1931magn}
S.~Chapman and V.~C.~A.~Ferraro.
"A new theory of magnetic storms,"
\href{http://dx.doi.org/10.1029/TE036i002p00077} {Terr. Magn.}, 36, pp.171-186, 1931.

\bibitem{shumlak2001evidence}
U.~Shumlak, R.~P.~Golingo, B.~A.~Nelson, and D.~J.~Den Hartog.
"Evidence of stabilization in the z-pinch,"
\href{https://doi.org/10.1103/PhysRevLett.87.205005}{Phys. Rev. Lett.}, 87(20):205005, 2001.

\bibitem{shumlak2003sheared}
U.~Shumlak, B.~A.~Nelson, R.~P.~Golingo, S.~L.~Jackson, E.~A.~Crawford, and D.~J.~Den Hartog.
"Sheared flow stabilization experiments in the zap flow z pinch,"
\href{http://dx.doi.org/10.1063/1.1558294}{Phys. Plasmas}, 10(5):1683--1690, 2003.

\bibitem{golingo2005formation}
R.~P.~Golingo, U.~Shumlak and B.~A.~Nelson.
"Formation of a sheared flow z pinch,"
\href{http://dx.doi.org/10.1063/1.1928249} {Phys. Plasmas}, 12(6):062505, 2005.

\bibitem{shumlak2006plasma}
U.~Shumlak, B.~A.~Nelson, and B.~Balick.
"Plasma jet studies via the flow z-pinch,"
\href{http://dx.doi.org/10.1007/978-1-4020-6055-7_8} {HEDLA}, pages 41--45. Springer, 2006.

\bibitem{shumlak2012sheared}
U.~Shumlak, J.~Chadney, R.~P.~Golingo, D.~J.~Den Hartog, M.~C.~Hughes, S.~D.~Knecht, W.~Lowrie, V.~S.~Lukin, B.~A.~Nelson, R.~J.~Oberto et~al.
"The sheared-flow stabilized z-pinch,"
\href{https://doi.org/10.13182/FST12-A13407} {Fusion Sci. Technol.}, 61(1T):119--124, 2012.

\bibitem{shumlak2017increasing}
U.~Shumlak, B.~A.~Nelson, EL.~Claveau, E.~G.~Forbes, R.~P.~Golingo, M.~C.~Hughes, R.~J.~Oberto, M.~P.~Ross, and T.~R.~Weber.
"Increasing plasma parameters using sheared flow stabilization of a z-pinch,"
\href{http://dx.doi.org/10.1063/1.4977468} {Phys. Plasmas}, 24(5):055702, 2017.

\bibitem{yamada1981quasistatic}
M.~Yamada, H.~P.~Furth, W.~Hsu, A.~Janos, S.~Jardin, M.~Okabayashi, J.~Sinnis, T.~H.~Stix and K.~Yamazaki.
"Quasistatic formation of the spheromak plasma configuration,"
\href{https://doi.org/10.1103/PhysRevLett.46.188} {Phys. Rev. Lett.}, 46(3), p.188, 1981.

\bibitem{jarboe1994review}
T.~R.~Jarboe.
"Review of spheromak research,"
\href{https://doi.org/10.1088/0741-3335/36/6/002} {Plasma Phys. Controlled Fusion}, 36(6), p.945, 1994.

\bibitem{treumann1997advanced}
R.~A.~Treumann and W.~Baumjohann.
Advanced space plasma physics (Vol. 30),
\href{https://doi.org/10.1142/p020} {London: Imperial College Press}, 1997.

\bibitem{dursi2008draping}
L.~J.~Dursi and C.~Pfrommer.
"Draping of Cluster Magnetic Fields over Bullets and Bubbles-Morphology and Dynamic Effects,"
\href{https://doi.org/10.1086/529371} {Astrophys. J.}, 677(2), p.993, 2008.

\bibitem{harris1962rayleigh}
E.~G.~Harris.
"Rayleigh-Taylor Instabilities of a Collapsing Cylindrical Shell in a Magnetic Field,"
\href{https://doi.org/10.1063/1.1724473} {Phys. Fluids}, 5(9), pp.1057-1062, 1962.

\bibitem{chandrasekhar1961hydrodynamic}
S.~Chandrasekhar.
Hydrodynamic and Hydrodynamic stability,
\href{https://doi.org/10.1063/1.3058072} {Clarendon Press}, 1961.

\end{thebibliography}
\end{document}